\documentclass[12pt]{article}
\usepackage{amsmath,amssymb,amsthm}
\usepackage[english]{babel}
\usepackage{graphicx, color, epsfig}
\usepackage{enumerate}
\usepackage{enumitem}
\usepackage{latexsym}
\usepackage{fancyhdr}
\usepackage{multirow}
\usepackage{geometry}
\usepackage[latin1]{inputenc}
\usepackage{hyperref}

\newcommand{\ds}{\displaystyle}
\newcommand{\vs}{\vspace}
\newcommand{\nt}{\nonumber}

\newcommand{\B}{\mathbf{B}}

\newcommand{\dd}{\mathrm{d}}

\newcommand{\E}{\mathbf{E}}
\newcommand{\e}{\mathrm{e}}

\newcommand{\h}{\mathrm{h}}

\newcommand{\ii}{\mathrm{i}}

\newcommand{\kk}{\mathbf{k}}

\newcommand{\nn}{\mathbf{n}}
\newcommand{\N}{\mathbb{N}}

\newcommand{\R}{\mathbb{R}}
\newcommand{\rr}{\mathbf{r}}

\renewcommand{\ss}{\mathbf{s}}

\newcommand{\Z}{\mathbb{Z}}

\newcommand{\zerog}{\textrm{\mathversion{bold}$0$}}

\newcommand{\ldc}{[\![}
\newcommand{\rdc}{]\!]}
\newcommand{\pr}{\partial}

\newcommand{\sign}{\textrm{sign}}

\newtheorem{prop}{Proposition}
\newtheorem{theo}{Theorem}
\newtheorem{subtheo}{Theorem}

\newtheorem{lem}{Lemma}
\newcommand{\cor}{{\bf Corollary.}~~}

\def\sqw{\hbox{\rlap{\leavevmode\raise.3ex\hbox{$\sqcap$}}$%
\sqcup$}}
\def\sqb{\hbox{\hskip5pt\vrule width4pt height6pt depth1.5pt%
\hskip1pt}}

\def\qed{\ifmmode\hbox{\hfill\sqb}\else{\ifhmode\unskip\fi%
\nobreak\hfil
\penalty50\hskip1em\null\nobreak\hfil\sqb
\parfillskip=0pt\finalhyphendemerits=0\endgraf}\fi}
\def\cqfd{\ifmmode\sqw\else{\ifhmode\unskip\fi\nobreak\hfil
\penalty50\hskip1em\null\nobreak\hfil\sqw
\parfillskip=0pt\finalhyphendemerits=0\endgraf}\fi}

\begin{document}
\renewcommand{\proofname}{{\bf Proof}}
  \begin{titlepage}
    \begin{center}
{\LARGE\bf\textsc{Lamb shift in non-degenerate energy
\\ \vs{.25cm}
level systems placed between two
\\ \vs{.25cm}
infinite parallel conducting plates}}
\ \\ \ \\ \ \\
{\bf Baptiste \textsc{Billaud}$^a$, T. T. \textsc{Truong}$^b$}
\ \\ \ \\ \ \\
$^a${\it Laboratoire de Math\'ematiques ``Analyse, G\'eometrie Mod\'elisation''
\\
Universit\'e de Cergy-Pontoise (CNRS UMR 8088), Saint-Martin 2
\\
2  avenue Adolphe Chauvin, F-95302 Cergy-Pontoise Cedex, France.}
\\
E-mail: {\tt bbillaud@u-cergy.fr}
\ \\ \ \\
$^b${\it Laboratoire de Physique Th\'eorique et Mod\'elisation
\\
Universit\'e de Cergy-Pontoise (CNRS UMR 8089), Saint-Martin 2
\\
2  avenue Adolphe Chauvin, F-95302 Cergy-Pontoise Cedex, France.}
\\
E-mail: {\tt truong@u-cergy.fr}
    \end{center} \ \\
    \begin{abstract}
\noindent
In this paper, the Lamb shift in systems with non-degenerate energy levels, put in the electromagnetic environment provided by two infinite parallel conducting plates, is analyzed. An explicit formula giving the relative Lamb shift (as compared to the standard one in vacuum) is derived for spherical semiconductor Quantum Dots (QD), \emph{via} a careful mathematical treatment of divergences in the calculations using the theory of distributions. This result settles a controversy between two different formulas existing in the current literature. Its sensitive dependence on the plates separation may be viewed as an indirect manifestation of the Lamb shift and may be used for the fine tuning of QD non-degenerate energy spectrum in some experimental contexts.
    \end{abstract} \ 
\end{titlepage}
\section{Introduction}
Both Lamb and Casimir effects were discovered in the late 1940s \cite{Lamb_1947, Casimir_1948}. They have been actively investigated ever since \cite{Eides_2001, Plunien_1985}, and have emerged as the strongest experimental supports for the quantization of the electromagnetic field.

In free space, the ground state of the quantized electromagnetic field is the siege of quantum fluctuations which generate the so-called zero-point energy and its observable effects. In hydrogen-like atoms, the $s$-levels are the only ones dressed by the electromagnetic field quantum fluctuations and thereby are split from other states, which should have been degenerate with them according to Dirac theory, giving rise the known Lamb shift \cite{Bethe_1947, Welton_1948}.

Quantum fluctuations also induce the so-called Casimir effect. In general, the summation of the zero-point energy fluctuations of the electromagnetic field yields a divergent ground state energy. In the absence of coupling to gravity, this divergence is usually subtracted off in an additive renormalization scheme. However, a careful analysis on its volume dependence reveals the occurrence of an observable force, known as Casimir force, which appears as an attractive force between two parallel conducting plates \cite{Casimir_1949}. Following an earlier and unfortunately not totally satisfactory attempt to detect this Casimir force \cite{Sparnaay_1958}, subsequent experiments turn out to be more convincing for plates separation distances in the ranges 0.6-6$\mu$m \cite{Lamoreaux_1997} and 0.1-0.9$\mu$m \cite{Mohideen_1998}. In these experiments, one of the plates has been replaced by a perfectly conducting sphere. This replacement allows to weaken the boundary effects due to the finite size of the plates, and thereby has facilitated its observation. From the theoretical point of view, great advances have also been done since Casimir original paper. The regularization arguments, underlying any calculation involving the Casimir effect, have been notably investigated in a mathematical rigourous way by several authors. Let us cite on this subject the influential contributions \cite{Elizalde_1994, Elizalde_1995, Kirsten_2002}. For a bibliography on the Casimir effect we refer the reader to articles published in \cite{Duplantier}.

In the context of Cavity Quantum Electrodynamics (CQED), our recent investigation on cavity-induced effects on atomic radiative properties of Quantum Dots (QDs) \cite{Brune_1996, Grynberg} has led us to wonder how much the Lamb shift of a system would change if it is put inside an environment provided by two parallel conducting plates, which shall be called, from now on, a {\it Casimir device}. The change of the atomic Lamb shift induced by a modification of the zero-point energy due to new electromagnetic field boundary conditions has been studied in \cite{Cheon_1988}. It has been also shown how hermitian conditions are sufficient to separate the contributions of vacuum fluctuations from those of self-radiation reaction to the energy level shifts \cite{Dalibard_1982}. The coupling of an electric dipole to a conducting surface through absorption and emission of its own radiative field is a well-known problem from a classical as well as from a quantum-mechanical point of view, see \emph{e.g.} \cite{Morawitz_1969, Holland_1984}. However, the understanding of the respective role of vacuum fluctuations and radiative field on the energy level shifts between two parallel plates seems to remain ambiguous \cite{Jhe_1991}. In this paper, to pursue investigations on the analogy between semiconductor QDs and real atoms \cite{Billaud_2010}, we address the problem of observability of the atomic-like Lamb shift in spherical semiconductor QDs. As the energy levels of an electron-hole pair confined in a spherical semiconductor QD are non-degenerate \cite{Efros_1982}, how would the Lamb shift in QDs shows up since the $s$- and $p$-levels are both shifted by the Lamb effect, as opposed to real atoms?

In section {\bf\ref{sec_2}}, the change in Lamb shift for a quantum system placed in a Casimir device is obtained by a careful mathematical treatment. In order to avoid masking the physics, the details of the calculational steps are given in the appendices. The net outcome is beneficial because it settles a controversy between two different expressions found in the literature, \emph{i.e.} in \cite{Cheon_1988} and in \cite{Jhe_1991}. Then in section {\bf\ref{sec_3}}, we apply the obtained results to the case of a spherical semiconductor QD, described in the formalism of the effective mass approximation (EMA) \cite{Brus_1984, Kayanuma_1988}. The explicit dependence of the additional Lamb shift on the separation distance of the plates in a Casimir device suggests a convenient way to modulate the QD energy spectrum for experimental purposes as well as to propose a way to "uncover" the Lamb shift in semiconductor QDs, at least in a so-called strong confinement regime and for a judiciously chosen semiconductor. A concluding section summarizes our main results in the last section.
\section{Effect of a Casimir device on the Lamb shift} \label{sec_2}
Inside two parallel perfectly conducting squared plates of linear size $L$ placed at a distance $d\ll L$ in vacuum, because the zero-point energy fluctuations are more important outside than inside the volume defined by the plates, a Casimir effect arises as an attractive force between the plates \cite{Casimir_1948}.

For an atom the phenomenological argument of Welton \cite{Welton_1948} shows that the Lamb shift of an energy level may be viewed as due to the particle position fluctuations induced by the zero-point energy of the unrestricted surrounding electromagnetic field. Consequently, when this surrounding electromagnetic environment is changed to that of a Casimir device, one should expect that the Lamb shift takes a different form and value for a given quantum state specified by a same set of quantum numbers. We argue that this difference in Lamb shifts may be exploited to \emph{reveal} the Lamb shift in systems with non-degenerate energy-levels.

Historically the Lamb shift in real atoms placed in a Casimir device has been calculated with the Bethe approach for a relativistic electron in \cite{Cheon_1988}, as well as for a non-relativistic electron in \cite{Jhe_1991}. They both predict an additional shift which depends on the separation distance $d$ between the plates and goes to zero in the limit of $d\rightarrow\infty$. However, there is a discrepancy in the leading contribution to the additional Lamb shift in these two predictions \cite{Cheon_1988, Jhe_1991}. For hydrogen-like atoms, it was shown in \cite{Cheon_1988} that this additional shift is inversely proportional to the separation distance $d$, and that it is the sum of a non-relativistic contribution, scaling as the atomic Rydberg energy $E^*$, and of a relativistic correction inversely proportional to the electronic bare mass $m_\e$ --- in units with $\hbar=c=1$. In the non-relativistic limit, for which the typical binding energy of the electron to the nucleus is negligible against its typical rest energy, {\it i.e.} if $E^*\ll m_\e$, only the non-relativistic contribution survives, so that the additional shift finally scales as $\propto\frac{E^*}d$, where the symbol ``\,$\propto$\,'' stands for ``proportional to''. This is in flagrant contradiction to the result of \cite{Jhe_1991}, which predicts, in the non-relativistic limit, that the leading contribution scales as $\propto\frac1{d^2}$. In this work, we show that the correct behavior goes effectively as $\propto\frac1{d^2}$, bringing to light the reasoning mistake in \cite{Cheon_1988}. We expect to reach the same conclusions in the Welton approach.
\subsection{Statement of the problem and assumptions}
To set up a framework for discussions, we shall use the Pauli-Fierz Hamiltonian in the Coulomb gauge $H_\mathrm{PF}$ \cite{Pauli_1938}, in which $H_0$ is the Hamiltonian of a non-relativistic spinless particle of mass $m^*$ and of charge $\pm qe$. The Hamiltonian $H_0$ is assumed to have eigenstates $|\nn\rangle$ with energy eigenvalues $E_\nn$, where $\nn$ is the related set of quantum numbers.

Following \cite{Balian_2004}, we first develop an efficient way to compute the standard Casimir force between the plates, which is formulated and handled in the framework of distribution theory.\footnote{The detailed mathematical steps are presented
in appendices {\bf\ref{appendix_A}} and {\bf\ref{appendix_B}}.} The result will be used
to obtain the density of states modified by the Casimir device.  The modification of the density of states induces the expected modification of the Lamb shift undergone by the energy levels of our system.

To this end, we evaluate the energy shift due to the vacuum fluctuations. However
the effect of the particle self-radiation reflected by the plates shall not be taken into account. We then fix a regime in which $d$ is sufficiently small to allow the emergence of
the Casimir effect between the plates, {\it i.e.} $d$ should be at most of the order
of magnitude of $\mu$m \cite{Mohideen_1998}. Furthermore, we assume that the typical wavelength, which scales as $\propto d$, due to the confinement between the Casimir plates, should be greater than the wavelength associated to an authorized radiative transition
between two energy levels of the particle described by the Hamiltonian $H_0$. According to \cite{Morawitz_1969, Holland_1984}, it is known that, in such a weak coupling regime, the coupling of a two-level quantum atom to itself through absorption and emission of dipolar radiation reflected by the Casimir plates is dominated by the coupling of this two-level atom to the electromagnetic field vacuum fluctuations. In a hydrogen-like atom, the associated Rydberg energy $E^*$ typically characterizes radiative transitions, their wavelength scaling as $\propto\frac1{E^*}$. Then, the weak coupling regime means that
  \begin{equation}
\kappa_d<\kappa^*, \label{condition}
  \end{equation}
where $\kappa_d=\frac\pi d$ is the ground state energy in the presence of the Casimir plates. 

For real atoms, the usual Bethe average excitation energy $\kappa^*\propto E^*$, used as IR cut-off in the Lamb effect  \cite{Bethe_1947, Welton_1948}, is surprisingly higher than the associated Rydberg energy $E^*$, although it represents the maximal excitation energy the atom could access \cite{Bethe_1950}.

In semiconductor QDs, the coupling of the quantum particle placed in a Casimir device to its own radiation field shall be also discarded. The validity of Eq. (\ref{condition}) for spherical semiconductor QDs will be examined in more details in section {\bf\ref{sec_3}}. As we shall see in this particular case, the Bethe cut-off $\kappa^*$ is of the same order of magnitude as the typical QD radiative transition energy. It is then also the natural candidate to depict the characteristic properties of the quantum system under study, as far as Lamb effect is concerned. Contrary to real atoms, the wavefunctions of an electron-hole pair confined in a QD, described by the standard effective mass approximation (EMA), are restricted to the region of space defined by the QD boundary surface \cite{Efros_1982, Brus_1984}. Then the probability for the electron, and to a lesser extent for the hole, to tunnel from the inside part of the QD to its outside surrounding is vanishingly small. Actually, even if the confinement potential exerted on the electron-hole trapped in the semiconductor QD is more appropriately represented by a finite potential step \cite{Thoai_1990}, the assumption of an infinite potential wall remains reasonable, since the tunnel effect probability remains exponentially small. Thus, by construction, the probability of interaction between a photon emitted by the QD inside the Casimir plates and the QD itself is negligible in comparison to the probability of its interaction with an excitation of the surrounding quantized electromagnetic field.
\subsection{Standard Casimir effect} \label{subsec_2_2}
As stated in \cite{Balian_2004}, the limit of perfectly conducting plates allows to consider the Casimir effect as only a manifestation of electromagnetic field vacuum fluctuations by uncoupling non-ambiguously the zero-point energy of the electromagnetic field from the surrounding matter. The electric and magnetic fields $\E$ and $\B$ are now supposed to satisfy the continuity boundary conditions on the plates
  $$
\E_\parallel=\B_\perp=\zerog,
  $$
where the indices ``$\parallel$'' and ``$\perp$'' stand for the tangential and orthogonal components of the fields with respect to the plates. A convenient way to describe the Casimir effect is to consider a rectangular box $\mathcal B$ of volume $V=L^2d$, built from the Casimir plates, as a waveguide running along a direction orthogonal to the plates, which shall be called the $z$-direction from now on. Then, the set of TM and TE modes forms a natural functional basis for the electromagnetic field.\footnote{For more details, one can refer to sections {\bf8.1} and {\bf8.2} of \cite{Jackson} In particular, there also exists a TEM mode, which is non-trivial only if the waveguide section is simply connected.} Denoting 
the tangential wave vector by $\kk_\parallel=\frac{2\pi}L\nn_\parallel$ and the orthogonal wave vector by $k_\perp=\kappa_dn_\perp$, being a natural cut-off wave number due to the finite size of the waveguide in the $z$-direction, this means that, except for modes for which $n_\perp=0$, each mode, defined by a wave-number $k_{\nn_\parallel n_\perp}=\sqrt{\kk^2_\parallel+(\kappa_dn_\perp)^2}$ and quantum numbers $(\nn_\parallel,n_\perp)\in\Z^2\times\N\smallsetminus\{0\}$,\footnote{The sets $\N\smallsetminus\{0\}$, $\Z\smallsetminus\{0\}$ and $\R\smallsetminus\{0\}$ are respectively the set of strictly positive integers, the set of non-zero relative integers and the set of non-zero real numbers.} has two possible polarizations. If periodic boundary conditions along the plates are also imposed, the zero-point energy of this electromagnetic field is then given by the divergent series
  $$
E(L,d)=\sum_{(\nn_\parallel,n_\perp)\in\Z^2\times\N\smallsetminus\{0\}}k_{\nn_\parallel n_\perp}+\sum_{\nn_\parallel\in\Z^2}\frac{k_{\nn_\parallel0_\perp}}2=\sum_{(\nn_\parallel,n_\perp)\in\Z^2\times\Z}\frac{k_{\nn_\parallel n_\perp}}2.
  $$
Reference \cite{Balian_2004} gives a prescription for regularizing this expression. Recalling that even if the plates are supposed to be perfectly conducting, any conducting material is transparent to radiation at sufficiently high frequencies. Then modes of arbitrary high frequencies do not actually contribute to the Casimir force between the plates.

This can be done by introducing a dimensionless cut-off function $\phi\!\left(\frac k{\kappa_\phi}\right)\!$, where $\kappa_\phi$ is a UV cut-off, in the expression of the zero-point energy $E(L,d)$, as follows
  \begin{equation}
E_\phi(L,d)=\sum_{(\nn_\parallel,n_\perp)\in\Z^2\times\Z}\frac{k_{\nn_\parallel n_\perp}}2\,\,\phi\!\!\left(\frac{k_{\nn_\parallel n_\perp}}{\kappa_\phi}\right)\!. \label{E_phi(L,d)_sum}
  \end{equation}
Without loss of generality, $\phi$ is assumed to verify $\phi(0)=1$, so that, in the limit of a perfect conductor, {\it i.e.} $k\ll \kappa_\phi$, each term appearing in the sum defining the regularized $E_\phi(L,d)$ goes to the term with the same quantum numbers of the sum defining non-regularized $E(L,d)$. It is very useful to suppose moreover that $\phi\in\mathcal S(\R)$, the Schwartz space of smooth and rapidly decreasing functions on $\R$.\footnote{For more details on distributions theory of the Lebesgue integral, one can respectively refer to \cite{Schwartz_Duistermaat_Strichartz}.} This assumption allows us to rigourously justify our calculations in the distribution sense. The details of the calculations are given in appendix {\bf\ref{appendix_B}}, and the zero-point energy of the electromagnetic field in presence of the Casimir device is found to be
  \begin{equation}
E_\phi(L,d)=V\!\int_{\R_+}\!\!\dd k~\!\frac{k^3}{2\pi^2}\,\,\phi\!\!\left(\frac k{\kappa_\phi}\right)\!+V\!\int_{\R_+}\!\!\dd k~\!\frac{\kappa_dk^2}{2\pi^3}g\!\! \left(\frac k{\kappa_d}\right)\!\phi\!\!\left(\frac k{\kappa_\phi}\right)\! \label{E_phi(L,d)}.
  \end{equation}
Recall that the standard Casimir energy between the two plates is
  $$
E_\mathrm{Casimir}(L,d)=-\frac{\pi^2}{720}\frac{L^2}{d^3}.
  $$
To evaluate $E_\phi(L,d)$, we need the explicit form the function $g(s)$ in Eq. (\ref{E_phi(L,d)}), which is given by the series
  $$
g(s)=\sum_{p\in\N\smallsetminus\{0\}}\frac{\sin(2\pi ps)}p,~~\forall s\in\R.
  $$
The computation of the function $g(s)$, which is quite involved and thoroughly explained in  appendix {\bf\ref{appendix_A}}, yields the expression $g(s)=\arctan(\tan\pi(\frac12-s))\,\chi_{\R\smallsetminus\Z}(s)$, for any $s\in\R$, where $\chi_\mathbb A(x)=\left\{
    \begin{array}{rl}
1 & \textrm{if}~~x\in\mathbb A
\\
0 & \textrm{otherwise}
    \end{array}
\right.\!\!$ is the so-called characteristic function of the subset $A\subseteq\R$. This result is crucial since it shows that the form of $g(s)$ as computed in appendix \textbf{A} of \cite{Cheon_1988} is incorrect.

The calculation of the zero-point energy of the electromagnetic field shows how its density of states is modified by the presence of the plates. This modification can be viewed as a perturbation of the free-space case, {\it i.e.} without Casimir device. More precisely, the first term of Eq. (\ref{E_phi(L,d)}) consists of the well-known contribution of the zero-point energy of the electromagnetic field in a region of free-space of volume $V$, because of its characteristic behavior as the third power of the mode eigenenergy $k$. The second term, which is the only one dependent on the separation distance $d$, is then the contribution due to the presence of the plates, which shall be denoted by $E_\mathrm{Casimir}^\phi(L,d)$. In particular, it is actually, as expected, independent of the regularization function $\phi$ in the limit $\frac{\kappa_d}{\kappa_\phi}\rightarrow0$. This is coherent with the fact that the Casimir energy is a physical quantity, as opposed to the first term, which is explicitly regularized by it. Therefore, the second term of Eq. (\ref{E_phi(L,d)}) suggests that the modification of the zero-point energy due to the Casimir device can be indeed entirely determined by a correction to the standard density of states $\rho(k)=\frac{k^2}{\pi^2}$ in free-space \cite{Itzykson}, given by
  \begin{equation}
\rho_d(k)=\frac{\kappa_dk}{\pi^3}g\!\!\left(\frac k{\kappa_d}\right)\!.
  \end{equation}
We insist on the fact that this function is not strictly a density, since it is not positive.
However, this should be considered as a perturbation to the case of the absence of the plates in the sense that, for a fixed mode of energy $k>0$ and in the limit $\frac k{\kappa_d}{}\rightarrow\infty$ of infinite separation distance between the plates, the correction $\rho_d(k)$ vanishes, and the free-space density of states $\rho(k)$ is recovered in this limit
  $$
\left|\frac{\rho_d(k)}{\rho(k)}\right|=\frac1\pi\frac{\kappa_d}k\left|g\!\!\left(\frac k{\kappa_d}\right)\right|\!\leq\frac{\kappa_d}{2k}\xrightarrow[\frac k{\kappa_d}{}\rightarrow\infty]{}0.
  $$
This insures that the usual free-space properties should be retrieved, when the separation distance becomes large. This is the reason why we will refer to the function $\rho_d(k)$ as a density by abuse of language. Furthermore, it is shown in appendix {\bf\ref{appendix_B}} that this computation method may be applied to any function $F(k)$ of the eigenmode energy $k$, when its mean value between the Casimir plates is to be evaluated. The total density of states between the two parallel plates is then the sum of the density of states $\rho(k)$ and a perturbation coming from the density of states $\rho_d(k)$, due to the plates. Therefore, to study the modification of a physical quantity due to the presence of the Casimir plates, we focus on the contribution due to the correction $\rho_d(k)$, given formally by the relation
  $$
\int\dd k~\!F(k)\rho_d(k)\phi\!\!\left(\frac k{\kappa_\phi}\right)\!=\frac{\kappa_d^{\gamma+2}}{\pi^3}\int\dd s\!~g(s)f^\beta(s)f_\gamma(s)\phi\!\!\left(\frac {\kappa_d}{\kappa_\phi}s\right)\!,
  $$
the free-space contribution of the standard density of states $\rho(k)$ being set aside. The functions $f^\beta(s)$ and $f_\gamma(s)$, as well as the indices $\beta$ and $\gamma$ are defined in appendix {\bf\ref{appendix_B}}. Intuitively they respectively contain the regular part of the function $F(k)$ and its IR divergent part, which must be appropriately regularized. This reasoning allows to identify directly the modification of the Lamb shift coming from the presence of the plates in both Bethe \cite{Bethe_1947} and Welton \cite{Welton_1948} approaches, as discussed in the coming subsections.
\subsection{Bethe et Welton approaches} \label{subsec_2_3}
As known, the Bethe approach to the Lamb effect is purely pertubative. The quantum second order time independent degenerate perturbation theory is applied to the Pauli-Fierz Hamiltonian $H_\mathrm{PF}$, where the electromagnetic field is treated in the weak field limit \cite{Bethe_1947}. Using renormalization arguments, in free-space, the Lamb shift undergone by any energy eigenstate $|\nn\rangle$ of the quantum system is found to be
  \begin{equation}
\Delta E_\nn=\frac\alpha{3\pi}\frac{q^2}{m^{*2}}\log\frac{m^*}{\kappa^*}\langle\nn|\nabla^2V(\rr)|\nn\rangle, \label{DeltaE_lamb}
  \end{equation}
where $m^*$ is used as a natural UV cut-off, since the assumption of non-relativistic particle is assumed here. Historically, this predicts a Lamb shift for the hydrogen atom $2s$-level, which is in excellent agreement with experimental values \cite{Bethe_1947, Bethe_1950}.

The Welton approach is also a perturbative approach, but has a more phenomenological aspect.
It has the merit of giving a physical picture of the origin of the Lamb effect.
More precisely, the Lamb shift is interpreted as a fluctuation effect on the particle position due to its interaction with the surrounding electromagnetic field. These fluctuations $\Delta\rr$ can be treated as a continuous random variable. Its probability density is a three-dimensional centered isotropic Gaussian distribution of variance $\langle(\Delta\rr)^2\rangle=\frac{2\alpha}\pi\frac{q^2}{m^{*2}}\log\frac{m^*}{\kappa^*}$, where $\kappa^*$ is the Bethe IR cut-off, and $m^*$ is used as a natural UV cut-off consistent with non-relativistic assumption, discarding fluctuation modes of order of the particle Compton wavelength \cite{Welton_1948}. The particle then moves in a new effective potential $\langle V(\rr+\Delta\rr)\rangle$, averaged on the fluctuation distribution. The first corrective term $\Delta V(\rr)$ in the fine structure constant $\alpha$ is precisely the term giving rise to the Lamb shift
  $$
\Delta E_\nn=\int\!\dd^3\rr\left|\langle\rr|\nn\rangle\right|^2\Delta V(\rr)=\frac\alpha{3\pi}\frac{q^2}{m^{*2}}\log\frac{m^*}{\kappa^*}\langle\nn|\nabla^2V(\rr)|\nn\rangle.
  $$
\subsection{Modification of the Lamb shift by a Casimir device}
Eq. (\ref{DeltaE_lamb}) is a regularized version of Eq. (5) in Bethe original paper \cite{Bethe_1947}, written in the formalism we have introduced previously, and in the non-relativistic limit $\kappa^*\ll m^*$
  $$
\Delta E^\mathrm{Bethe}_\nn=\frac{\pi\alpha}3\frac{q^2}{m^{*2}}\!~\langle\nn|\nabla^2V(\rr)|\nn\rangle\int_0^{m^*}
\frac{\dd k}{k+\kappa^*}\frac{\rho(k)}{k^2}.
  $$
Eq. (\ref{DeltaE_lamb}) is also a regularized version of Eq. (3) of Welton original paper \cite{Welton_1948}, in the non-relativistic limit $\kappa^*\ll m^*$,
  $$
\Delta E^\mathrm{Welton}_\nn=\frac{\pi\alpha}3\frac{q^2}{m^{*2}}\!~\langle\nn|\nabla^2V(\rr)|\nn\rangle\int_{\kappa^*}^{m^*}
\frac{\dd k}{k^3}\rho(k).
  $$
Therefore, invoking the remarks made in subsection {\bf\ref{subsec_2_2}}, in both cases,
the correction to the Lamb shift due to the presence of the Casimir device should be evaluated by replacing the density of states $\rho(k)$ in the absence of the Casimir device by the corrective term $\rho_d(k)$, and the divergent integral is regularized following the prescriptions of appendix {\bf\ref{appendix_B}} with a  function $\phi\in\mathcal S(\R)$. This reasoning is a shortcut to adapt Bethe or Welton original arguments to this new framework. Let us insist on the fact that, in free-space, both methods are equivalent and give the same results. In particular, they prescribe the same way to regularized the IR divergence of the previous integral. This will not be so inside a Casimir surrounding. Considerations on the Lamb shift in the non-relativistic limit provide a natural UV cut-off when needed, $m^*$ being the mass of the particle under study, {\it i.e.} here we will set $\kappa_\phi=m^*$.

Now in the Bethe approach, we recognize that $f^\beta_{\eta^*}(s)\propto\frac{\kappa^{-1}_d}{s+\eta^*}$, $f_\gamma(s)=\frac1s$, $\beta=1$ and $\gamma=-1$, where $\eta^*=\frac{\kappa^*}{\kappa_d}$. But in the  Welton approach, we see that $f(s)\propto1$, $f_\gamma(s)=\frac1{s^2}$, $\beta=0$ and $\gamma=-2$. In both cases, we should turn to theorem {\bf\ref{theorem_1}} (but not proposition {\bf\ref{proposition_6}}) of appendix {\bf\ref{appendix_B}} and more particularly to theorem {\bf\bf\ref{subtheorem_2a}}, to see that a second IR cut-off $\kappa_*=\eta_*\kappa_d$ is needed here, because $\gamma$ takes negative integer values.

As far as Lamb effect is concerned, the Bethe IR cut-off $\kappa^*$ is the perfect choice. Thus we set $\kappa_*=\kappa^*$. Moreover, thanks to the weak field coupling $\kappa^*>\kappa_d$, the regularization parameter $\eta^*$, defined by $\kappa^*$, verifies $\eta^*=\eta_*>1$. Theorem {\bf\bf\ref{subtheorem_2a}} then gives a Taylor expansion in power series of $\frac1{\eta_*}$ of the modification to the Lamb shift due to the Casimir plates for each prescription in order to regularize the wanted integral.

After having performed the Taylor expansion of the function $s\longmapsto f^\beta(s)f_\gamma(s_{\eta^*})$, we get for any $r\in\N\smallsetminus\{0\}$
  $$
\frac{\Delta E_\mathrm{Casimir}^{\mathrm{Bethe},\nn}(d)}{\Delta E_\nn}=\log^{-1}\frac{m^*}{\kappa^*}\!\left\{\sum_{n=1}^r\frac{b_{2n}}{2n(2n-1)\eta_*^{2n}}\sum_{p=0}^{n-1}\frac{(-1)^p\Gamma(p+\frac12)}{p!\Gamma(\frac12)}+O\!\!\left(\frac1{\eta_*^{2(r+1)}}\right)\!\right\}\!,
  $$
where $\Gamma(x)$ is the standard Gamma function, the real numbers $b_{2n}$ are the non-vanishing Bernoulli numbers,\footnote{For more details on special functions and Bernoulli numbers, one can refer to \cite{Abramowitz, Prudnikov, Gradshteyn}.} and
  $$
\frac{\Delta E_\mathrm{Casimir}^{\mathrm{Welton},\nn}(d)}{\Delta E_\nn}=\log^{-1}\frac{m^*}{\kappa^*}\!\left\{\sum_{n=1}^r\frac{|b_{2n}|}{2n(2n-1)\eta_*^{2n}}+O\!\!\left(\frac1{\eta_*^{2(r+1)}}\right)\!\right\}\!.
  $$
These expansions do not converge as power series. Actually they are strongly divergent,
and should be understood as asymptotic series.\footnote{\label{footnote_2} While the expansions $\Delta E_\mathrm{Casimir}^{\mathrm{Bethe},\nn}(d)$ or $\Delta E_\mathrm{Casimir}^{\mathrm{Welton},\nn}(d)$ do not converge, it is possible to recognize the so-called Stirling series, which is known to give the asymptotic behavior of the Gamma function in the neighborhood of $|z|\rightarrow\infty$. More precisely, from Eq. {\bf 6.1.42} \cite{Abramowitz}, in the neighborhood of $z\rightarrow\infty$, with $|\arg z|<\pi$, for any $r\in\N\smallsetminus\{0\}$
  $$
\sum_{n=1}^r\frac{b_{2n}}{2n(2n-1)z^{2n}}=\frac{\log\Gamma(z)-(z-\frac12)\log z+z-\frac12\log2\pi}z-R_{r+1}(z),
  $$
where the asymptotic behavior of the remainder $R_r(z)$ is given by
  $$
|R_r(z)|\leq\frac{|b_{2r}|}{2r(2r-1)|z|^{2r}}.
  $$
Moreover, from Eq. {\bf5.2.11.16} \cite{Prudnikov}, Abel theorem for series \cite{Gourdon_Whittaker} and Stirling formula {\bf6.1.39} \cite{Abramowitz}, we deduce, for any $n\in\N\smallsetminus\{0\}$
  $$
\left|\sum_{p=0}^{n-1}\frac{(-1)^p\Gamma(p+\frac12)}{p!\Gamma(\frac12)}-\frac1{\sqrt2}\right|\leq\frac{\Gamma(n+\frac12)}{n!\Gamma(\frac12)}\sim\frac1{\sqrt n}~~\Rightarrow~~\sum_{p=0}^{n-1}\frac{(-1)^p\Gamma(p+\frac12)}{p!\Gamma(\frac12)}=\frac1{\sqrt2}+O\!\!\left(\frac1{\sqrt n}\right)\!\!,
  $$
which implies in particular that
  $$
\frac{\Delta E_\mathrm{Casimir}^{\mathrm{Bethe},\nn}(d)}{\Delta E_\nn}<\frac{\Delta E_\mathrm{Casimir}^{\mathrm{Welton},\nn}(d)}{\Delta E_\nn}.
  $$} Asymptotic series are typical objets one encounters in Quantum Electrodynamics.
For example when applying the usual perturbation procedure or computing Feynmann diagrams a in the fine structure constant $\alpha$, the obtained series makes sense at all fixed order of the perturbation parameter $\alpha$, but do not asymptotically converge as a power series.

Let us observe that$^\mathrm{\ref{footnote_2}}$
  $$
\frac{\Delta E_\mathrm{Casimir}^{\mathrm{Bethe},\nn}(d)}{\Delta E_\nn}<\frac{\Delta E_\mathrm{Casimir}^{\mathrm{Welton},\nn}(d)}{\Delta E_\nn}.
  $$
This result may be expected, since the Bethe approach integrates renormalization arguments
whereas the Welton approach is purely phenomenological. In the Bethe approach,
the logarithmic IR divergence is indeed regularized by construction, while in the Welton approach it is regularized {\it ad hoc}. Then, when the usual density of states $\rho(k)$ is replaced by the correction due to the Casimir plates, this generates a new IR logarithmic divergence in the first case, and a divergence, scaling as $\propto\frac1k$ in the second. This explains why the modification is less important in the Bethe approach than in the Welton approach.

While the expressions for the modification to the Lamb shift $\Delta E_\mathrm{Casimir}^{\mathrm{Bethe},\nn}(d)$ and $\Delta E_\mathrm{Casimir}^{\mathrm{Welton},\nn}(d)$ differ, they exhibit the same first order behavior, which scales as $\propto\frac1{\eta_*^{2}}$. The difference between the predictions made by Bethe and Welton approaches indeed becomes significant either when the two IR cut-offs $\kappa_d$ and $\kappa^*$ are close enough, or when we consider a sufficiently large order in $\frac1{\eta^*}$, such that the expansion begins to diverge. We will therefore focus on value of $r\in\N\smallsetminus\{0\}$ such that $\frac{|b_{2r}|}{2r(2r-1)\eta_*^{2r}}<1$. Under these assumptions, orders higher than two in $\frac1{\eta_*^2}$ are negligible in comparison to the shared first order, which is then sufficient in the weak coupling limit $\kappa^*>\kappa_d$. Finally, for $\langle\nn|\nabla^2V(\rr)|\nn\rangle\neq0$, {\it i.e.} when the energy level under study actually undergoes the Lamb effect, we may retrieve the relative modification of the Lamb shift due to the Casimir plates as
  \begin{equation}
\frac{\Delta E_\mathrm{Casimir}^\nn(d)}{\Delta E_\nn}=\frac1{12}\left(\frac{\kappa_d}{\kappa^*}\right)^{\!\!2}\left\{1+O\!\!\left[\left(\frac{\kappa_d}{\kappa^*}\right)^{\!\!2}\right]\right\}\log^{-1}\frac{m^*}{\kappa^*}\geq0. \label{DeltaE_Casimir(d)}
  \end{equation}
Because of the positive sign, this always gives rise to a enhancement of the Lamb shift.
This result calls for a physical explanation. When the separation distance $d$ decreases,
the amplitude of the electromagnetic modes inside the Casimir plates increases, while their number remains constant. This leads to the reinforcement of the interaction of the quantum system with the quantized electromagnetic field, implying a strengthening of the Lamb effect. Moreover, this relative enhancement does not depend on the quantum state under consideration. However, it shows an explicit a competition between the different scales of energies $\kappa_d<\kappa^*\ll m^*$, and consequently between the characteristic lengths of the problem, as we shall see in the next section. This expression is in agreement with \cite{Jhe_1991}, because up to the smallest order in the dimensionless IR cut-off $\eta^*$, the correction $\Delta E_\mathrm{Casimir}^\nn(d)$ scales as $\propto\frac1{\eta_*^{2}}\propto\frac1{d^2}$. In a regime where the quantum system does not interact with its own radiative field, our approach allows to compute explicitly the proportionality factor. In particular, it has been possible to factorize the mean value of the Laplace operator of the potential $\langle\nn|\nabla^2V(\rr)|\nn\rangle$, which is considered as a feature of the Lamb effect according to Welton approach.
%
%
%
%
\section{Observability of the Lamb shift in spherical semiconductor QDs} \label{sec_3}
As it is known, the experimental observability of the Lamb shift in hydrogen-like atom is due to the $s$- and $p$-level degeneracy, when the principal quantum number is $n\geq 2$, in the absence of interaction with the quantized electromagnetic field. The Lamb shift arises as a separation of the $ns$-spectral band from $np$-spectral band, when the interaction with the quantized surrounding electromagnetic field is taken into account. So, how would the Lamb shift of an energy level show up in quantum systems displaying no spectral band degeneracy such as a QD? In such systems each non-degenerate energy level is dressed by the quantum zero-point fluctuations of the electromagnetic field, forbidding the detection of the corresponding bare energy level. Here we can see that Eq. (\ref{DeltaE_Casimir(d)}) may be used to \emph{label} the Lamb shift in semiconducting QDs (calculated for the first time in \cite{Billaud_2010}) by each value of $d$.

The model we use to obtain this Lamb shift in QDs is an improved version of the standard EMA,
in which a pseudo-potential is introduced to partly remove the usual divergence of the QD ground state energy for small radii. One electron and one hole, moving with their standard effective masses $m^*_{\e,\h}$ in a semiconductor substrate, are confined by an infinite spherical potential well of radius $R$ and interact with each other through the Coulomb potential. The common approach to treat the interplay of the Coulomb interaction of the electron-hole pair, which scales as $\propto\frac1R$, and the quantum confinement, which scales as $\propto\frac1{R^2}$ is to use a variational procedure, for which two regimes of electron-hole pair should be singled out according to the ratio of the Bohr radius $a^*=\frac\kappa{e^2\mu}$ of the exciton, $\mu$ being its reduced mass, to the QD radius $R$. First, in the strong confinement regime, valid for a QD radius $R\leq2a^*$, the confinement potential sufficiently affects the relative electron-hole motion, so that the interactive electron-hole pair states should then consists of uncorrelated electron and hole states. In the weak confinement regime, valid for a QD radius $R\geq4a^*$, the electron-hole relative motion is left almost unchanged by the confinement potential, so that excitonic binding states appear, as if the electron-hole pair has not been confined. However, the exciton has to be treated as a confined quasi-particle of total mass $M$, and its center-of-mass motion should then be quantized.

\begin{figure}
\caption{Modification of the Lamb shift in spherical InAs microcrystals for $d=1\mu$m (---), $0.5\mu$m (--$\!~$--), $0.25\mu$m (--$\!~\cdot~\!$--) or $0.1\mu$m (--$\!~\cdot~\!\!\!~\cdot\!~$--) as a fonction of the QD radius $R$.} \label{figure_1}
    \begin{center}
\input{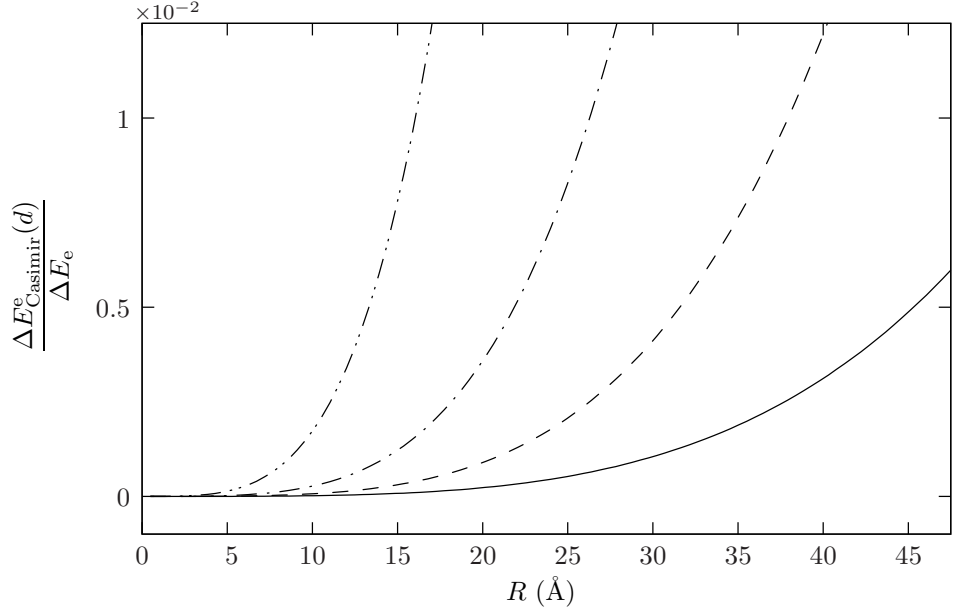}%
    \end{center}
  \end{figure}

For this simple model, in the strong confinement regime, it is proven that the predicted Lamb shift, in QDs of size experimentally synthetized and used, is of the same order of magnitude as the one in hydrogen-like atoms, at least for judiciously chosen semiconductors, such as for example InAs or GaAs, and then seems to be observable. Since this is the relative modification of the Lamb shift due to the presence of the Casimir device, the method of calculation of the Lamb shift in QDs \cite{Billaud_2010} is not of interest. However, this provides the Bethe IR cut-offs $\kappa^*_{\e,\h}=\frac{7\pi^2}{12m^*_{\e,\h}R^2}$ and $\kappa^*=\kappa^*_\e+\kappa^*_\h=\frac{7\pi^2}{12\mu R^2}$ respectively for the electron, the hole and the exciton. Let us introduce the electron and the hole reduced Compton wavelengths $\lambda^*_{\e,\h}=\frac1{m^*_{\e,\h}}$ and the radius $R^*_{\e,\h}=\frac\pi2\sqrt{\frac73}\lambda^*_{\e,\h}$. These are interpreted as the lower bound for the QD radius, allowing fluctuations of the charge carrier, according to the Welton approach, to be confined inside the QD. Then, it is supposed that $R^*_{\e,\h}\leq R\leq d$, with the additional constraint $\kappa_d<\kappa^*_{\e,\h}(R)$ given by Eq. (\ref{condition}). In this context, we can deduce from Eq. (\ref{DeltaE_Casimir(d)}) the relative modification of the Lamb effect undergone by a semiconducting QD placed in a Casimir device
  \begin{equation}
\frac{\Delta E_\mathrm{Casimir}^{\e,\h}(d)}{\Delta E_{\e,\h}}=\frac1{14}\!\left(\frac{R^2}{R^*_{\e,\h}d}\right)^{\!\!2}\!\!\left\{1+O\!\!\left[\left(\frac{R^2}{R^*_{\e,\h}d}\right)^{\!\!2}\right]\right\}\!\log^{-1}\frac R{R^*_{\e,\h}},~~~~\textrm{valid for}~~\frac{R^2}{R^*_{\e,\h}d}<\frac12\sqrt{\frac73}. \label{DeltaE^*_Casimir(d)}
  \end{equation}
Then, there is a competition between the dimensionless ratios $\frac R{R^*_{\e,\h}}\geq1$ and $\frac Rd\leq1$, characterizing the problem under study in the strong confinement regime, which is characterized in turns by the ratio $\frac R{a^*}\leq2$. Figure \ref{figure_1} shows the behavior of this modification inside InAs QDs as a function of the radius for several values of the separation distance $d$. Only the electronic contribution to the Lamb shift is represented on this figure, because, in InAs, either the electronic and hole contributions to the Lamb shift are almost equal (light hole, $\frac{m^*_\h}{m_\e}\approx\frac{m^*_\e}{m_\e}\approx0.026$), the effective masses being themselves almost identical, or the hole contribution is negligible against the electronic one (heavy hole, $\frac{m^*_\h}{m_\e}\approx0.41$).

Figure \ref{figure_1} indicates that if the separation distance $d$ is chosen to be 0.1$\mu$m, the experimental observation of the Casimir effect is possible. The modification of the electron-hole Lamb shift between the Casimir plates is of about 1\% in spherical InAs QDs of radius in the range of 10-15nm, which is of reasonable experimental size \cite{Banin_1998}. It is possible to enhance the amplitude of this modification by reducing the separation distance $d$ to the order of a few tenth parts of a $\mu$m. But the radius $R$ of the QD should be also reduced accordingly to satisfy the weak coupling regime condition. This effect alone almost leads to a modification of the Lamb shift in free-space by about 5\% of its value, which seems significant enough to be observable with the precision of nowadays experiments on Casimir effect.

\begin{figure}
\caption{Modification of the Lamb shift in spherical InAs microcrystals for $R=15$\AA~as a function of the plates separation $d$.} \label{figure_2}
    \begin{center}
\input{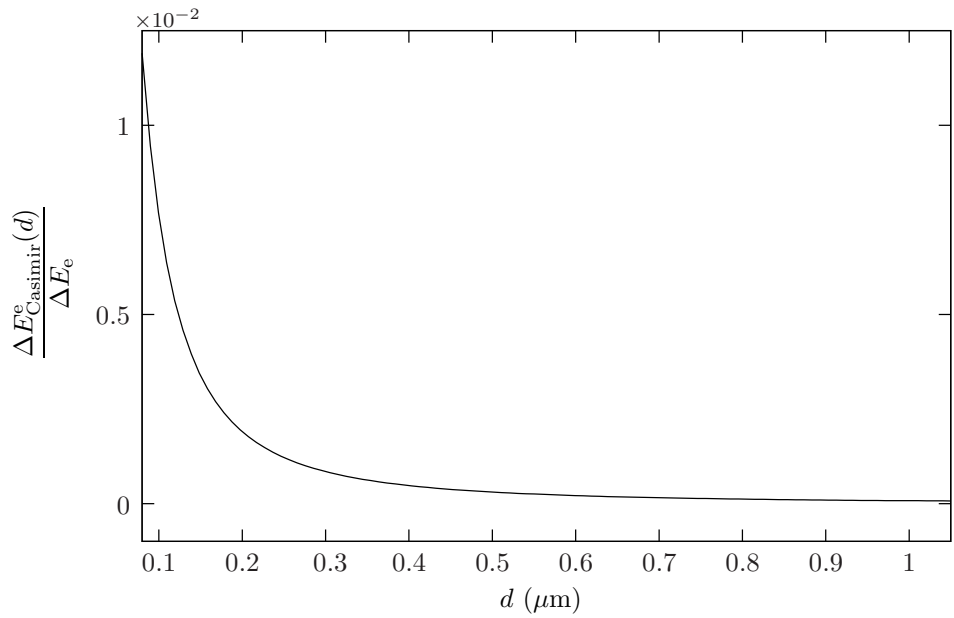}%
    \end{center}
  \end{figure}

Figure \ref{figure_2} illustrates for a given QD radius the modification of the Lamb shift as a function of the Casimir plates separation distance $d$. This suggests, within a certain range of $d$, the possibility of adjusting the energy spectrum of a QD when necessary in some experimental context. Alternatively may also consider the curve in figure \ref{figure_2} as illustrating the concrete manifestation of the Lamb shift in non-degenerate energy level systems.

There finally exists other corrective effects, such as the reflectivity of the metal used, the roughness of the surfaces of plates and sphere, and the finite temperature, which has an impact on the Casimir effect, as described in \cite{Mohideen_1998}. By the same kind of reasoning as above, it seems also possible to account for them in our description. Since these are corrections scaling as $\propto\frac1{d^2}$ to the standard Casimir force, it is sufficient to consider the two first terms in Eqs.  (\ref{DeltaE_Casimir(d)}) or (\ref{DeltaE^*_Casimir(d)}), which are the dominant terms of 
$\frac{\Delta E_\mathrm{Casimir}^{\e,\h}(d)}{\Delta E_{\e,\h}}$ and its first correction, also scaling as $\propto\frac1{d^4}$, is to be considered.
\section{Conclusion} \label{sec_4}
In this work, we have developed a comprehensive computation method to obtain mathematically rigorous results on the Lamb shift for non degenerate energy levels in semiconductor spherical quantum dots placed in a Casimir device. A deviation from the Lamb shift in vacuum, predicted in \cite{Billaud_2010}, is revealed. The explicit formula giving this deviation suggests the possibility to  fine tune the spectrum of a quantum dot for experimental purposes, by varying the separation distance between the Casimir plates, provided that such micro-mechanical operations are realizable. Moreover, the energy shift order of magnitude, at least for judiciously chosen semiconductor and QD sizes, seems to be sufficient to be checked experimentally in the present state of technology. Other possible uses of the Lamb shift in non-degenerate systems may not be totally out of reach.
\appendix
\section{On the evaluation of the sum of a trigonometric series} \label{appendix_A}
In this appendix, the function $g$ appearing in Eq. (\ref{E_phi(L,d)}) and defined by the series
  $$
\sum_{p\in\N\smallsetminus\{0\}}\frac{\sin(2\pi ps)}p,~~\forall s\in\R, 
  $$
is studied and its value calculated.

It is well-known that the previous series is a Fourier series on $\R$, which represents the function $g:s\longmapsto\arctan(\tan\pi(\frac12-s))\,\chi_{\R\smallsetminus\Z}(s)$. This function is odd and 1-periodic on $\R$, and is equal to $g(s)=\arctan(\tan\pi(\frac12-s))=\pi(\frac12-s)$ for $s\in]0,1[$. Since the function $g$ is piecewise continuous, we deduce from Dirichlet theorem \cite{Gourdon_Whittaker} that the Fourier series defining $g(s)$ is point-wise convergent on $\R$, and that
  \begin{equation}
g(s)=\sum_{p\in\N\smallsetminus\{0\}}\frac{\sin(2\pi ps)}p,~~\forall s\in\R. \label{g(s)}
  \end{equation}
The error made in \cite{Cheon_1988} is the statement that $g(s)=\pi(\frac12-s)$ for all $s\in\R$, from Eq. (A14) in appendix {\bf A} of this reference.
But, as seen above, the function $g$ is neither odd (because $g(0)\neq0$) nor periodic.

In the following, we establish the validity of Eq. (\ref{g(s)}) using the Euler-Maclaurin formula, as proposed in \cite{Cheon_1988}. The proof involves tricky points of regularization theory of series and distribution theory.
\subsection{Notations and convergences} \label{appendix_A1}
For any fixed $\epsilon>0$ and any $s\in\R$, consider the functions
  $$
g_\epsilon(t,s)=\frac{\sin(2\pi st)}t\e^{-\epsilon t^2}~~\textrm{and}~~g_0(t,s)=\frac{\sin(2\pi st)}t,~~\forall t\in\R.
  $$
From now on, we will denote \vs{-.2cm}
  \begin{itemize}
\item $g_\epsilon(t,\cdot):s\longmapsto g_\epsilon(t,s)$, for any $\epsilon$ and any $t$, for which it makes sense; \vs{-.2cm}
\item $g_\epsilon(\cdot,s):t\longmapsto g_\epsilon(t,s)$, for any $\epsilon$ and any $s$, for which it makes sense; \vs{-.2cm}
\item $g_\cdot(t,s):\epsilon\longmapsto g_\epsilon(t,s)$, for any $t$ and any $s$, for which it makes sense; \vs{-.2cm}
\item the usual supremum norm on any subset $\mathbb A\subseteq\R$ by $\lVert\cdot\rVert_\infty^\mathbb A=\sup_\mathbb A|\cdot|$. \vs{-.2cm}
  \end{itemize}
These notations will be extended as soon as needed.
\ \\
\par With these notations, we immediatly verify that $g_\epsilon(\cdot,s)\xrightarrow[\epsilon\rightarrow0^+]{}g_0(\cdot,s)$, point-wise on $\R$.

Let $s\in\R\smallsetminus\{0\}$, from Eqs. {\bf2.5.3.12} and {\bf2.5.36.6} \cite{Prudnikov}, we deduce
  $$
\int_{\R_+}\!\!\dd t~\!g_\epsilon(t,s)=\frac\pi2\,\sign(s)\,\mathrm{erf}\!\left(\frac\pi{\sqrt\epsilon}|s|\right)\xrightarrow[\epsilon\rightarrow0^+]{}\frac\pi2\,\sign(s)=\int_{\R_+}\!\!\dd t~\!g_0(t,s),
  $$
where $\sign=(\chi_{\R_+}-\chi_{\R_-})$ is the sign-function on $\R$, and $\mathrm{erf}$: $x\longmapsto\frac2{\sqrt\pi}\int_0^x\dd t~\!\e^{-t^2}$ is the error-function on $\R$, which verifies $\mathrm{erf}(x)\xrightarrow[x\rightarrow\infty]{}1$ and $|\mathrm{erf}|\leq1$ on $\R$. When $s=0$, it is obvious that $g_\epsilon(0,\cdot)=g_0(0,\cdot)=0$ on $\R$, and the previous expression is trivially satisfied.

Consider now the two point-wise convergent series on $\R$ defined by
  $$
g_\epsilon(s)=\sum_{p\in\N\smallsetminus\{0\}}g_\epsilon(p,s)=\sum_{p\in\N\smallsetminus\{0\}}\frac{\sin(2\pi ps)}p\e^{-\epsilon p^2}
  $$
and
  $$
g_0(s)=\sum_{p\in\N\smallsetminus\{0\}}g_0(p,s)=\sum_{p\in\N\smallsetminus\{0\}}\frac{\sin(2\pi ps)}p.
  $$

We first prove the following proposition.

  \begin{prop}~ \label{proposition_1}

$g_\epsilon\xrightarrow[\epsilon\rightarrow0^+]{}g_0$, point-wise on $\R$.
    \begin{proof}
For any $\epsilon\geq0$, $g_\epsilon$ is odd and 1-periodic on $\R$, so that it is sufficient to restrict the study of the point-wise convergence to $[0,\frac12]$.

For any $\epsilon\geq0$ and for $s\in\{0,\frac12\}$, we have $g_\epsilon(s)=0$.

Let $\epsilon>0$ and $s\in]0,\frac12[$.

The series $\sum g_\epsilon(p,s)$ are absolutely convergent, because $|g_\epsilon(p,s)|\leq\e^{-\epsilon p}$, $\forall p\in\N\smallsetminus\{0\}$, where $\e^{-\epsilon}\in\!~]0,1[$.

And, the series $\sum g_0(p,s)$ are semi-convergent thanks to Abel theorem \cite{Gourdon_Whittaker}. 

Abel theorem also yields, for any $N\in\N\smallsetminus\{0\}$, $\left\lVert\sum_{p\geq N}g_\cdot(p,s)\right\rVert_\infty^{\R_+}\leq\frac1{N\sin\pi s}\xrightarrow[N\rightarrow\infty]{}0$. This implies that the series $\epsilon\longmapsto\sum g_\epsilon(p,s)$ are uniformly convergent on $\R_+$.

Then, $\epsilon\longmapsto g_\epsilon(s)\in\mathcal C^0(\R_+)$, and in particular $g_\epsilon(s)\xrightarrow[\epsilon\rightarrow0^+]{}g_0(s)$.

By parity and periodicity extensions, one obtains finally $g_\epsilon\xrightarrow[\epsilon\rightarrow0^+]{}g_0$, point-wise on $\R$.
    \end{proof}
  \end{prop}
\ \\
\par As we shall see, this point-wise convergence is not sufficient, and we would rather have a convergence of $g_\epsilon$ to $g_0$ in the limit $\epsilon\rightarrow0^+$ in the space of tempered distributions $\mathcal S'(\R)$, denoted as $g_\epsilon\xrightarrow[\epsilon\rightarrow0^+]{}g_0$ in $\mathcal S'(\R)$.\footnote{For more details on the theory of the Lebesgue integral, one can refer to \cite{Farault_Malliavin}.}

The difficulty lies in the fact that the previous inequality $\left\lVert g_\cdot(s)\right\rVert_\infty^{\R_+}\leq\frac1{|\sin\pi s|}$, valid for any $s\in\R\smallsetminus\Z$ and then almost surely on $\R$, forbids the use of dominated convergence theorem, because $s\longmapsto\frac1{|\sin\pi s|}\notin\mathcal L^1_\mathrm{loc}(\R,\dd s)=\{h:\R\longrightarrow\R\!~\big|\!~\int_\mathbb K\dd s\!~|h(s)|<\infty,~\forall\mathbb K\subset\R~\textrm{compact}\}$. Therefore, it is not possible to directly deduce that $g_\epsilon\xrightarrow[\epsilon\rightarrow0^+]{}g_0$ in the space of distributions $\mathcal D'(\R)$. However, we shall prove in proposition {\bf\ref{proposition_4}} that there exists a constant $C>0$ such that $\lVert g_\cdot(s)\rVert_\infty^{[0,1]}\leq C$, almost surely on $[0,\frac12]$, which solves the problem.\footnote{As explained in introduction to this section, let us insist on the fact that we have adopted the notation $g_\cdot(s):\epsilon\longmapsto g_\epsilon(s)$.}

We now prove proposition 2. \ \\
  \begin{prop}~ \label{proposition_2}

$g_\epsilon\xrightarrow[\epsilon\rightarrow0^+]{}g_0$ in $\mathcal S'(\R)$.
    \begin{proof}
By parity and periodicity and proposition {\bf\ref{proposition_4}}, $\lVert g_\cdot(s)\rVert_\infty^{[0,1]}\leq C$, almost surely on $\R$, then $C\phi\in\mathcal L^1(\R,\dd s)=\{h:\R_+\longrightarrow\R\!~\big|\!~\int_{\R_+}\dd s\!~|h(s)|<\infty\}$, for any test function $\phi\in\mathcal S(\R)$. Hence, the functions $g_\epsilon$ are tempered distributions on $\R$, for any $\epsilon\geq0$, and the dominated convergence theorem applies and leads to the expected convergence
  $$
\int_\R\dd s\!~g_\epsilon(s)\phi(s)\xrightarrow[\epsilon\rightarrow0+]{}\int_\R\dd s\!~g_0(s)\phi(s),~~\forall \phi\in\mathcal S(\R)~~\Leftrightarrow~~g_\epsilon\xrightarrow[\epsilon_\mathrm{loc}0+]{}g_0~\textrm{in}~\mathcal S'(\R).
  $$
    \end{proof}
  \end{prop}
\subsection{Euler-Maclaurin formula and consequences} \label{appendix_A2}
The Euler-Maclaurin summation formula is an important tool in analysis. It provides an estimation of the sum $\sum_{p=0}^Nh(p)$ by the integral $\int_{[0,N]}\dd t~\!h(t)$, $h$ being a sufficiently regular function on $[0,N]$ with $N\in\N$. Let us assume, for convenience, that $h\in\mathcal C^\infty(\R)$, then for any $N\in\N$ and any $r\in\N\smallsetminus\{0\}$, we have
  $$
\sum_{p=0}^Nh(p)=\int_{[0,N]}\!\!\dd t~\!h(t)+\frac{h(N)+h(0)}2+\sum_{n=1}^r\frac{b_{2n}}{(2n)!}\left\{h^{(2n-1)}(N)-h^{(2n-1)}(0)\right\}+R_r^N(h),
  $$
where the remainder $R_r^N(h)$ is expressed as
  $$
R_r^N(h)=-\int_{[0,N]}\!\!\dd t~\!\frac{\widetilde B_{2r}(t)}{(2r)!}h^{(2r)}(t).
  $$
Here, the function $\widetilde B_{2n}$ is the unique 1-periodic function which coincide on $[0,1]$ with the Bernoulli polynomial $B_{2n}$, and the real numbers $\{b_{2n}\}_{n\in\N\smallsetminus\{0\}}$ are the non-vanishing Bernoulli numbers, defined by $b_{2n}=B_{2n}(0)=B_{2n}(1)$, for any $n\in\N\smallsetminus\{0\}$ \cite{Abramowitz}. A useful Fourier series representation of $\widetilde B_{2r}(t)$ is ({\it cf.} Eq. {\bf6.22} p. 1032 \cite{Gradshteyn})
  $$
\widetilde B_{2r}(t)=2(-1)^{r-1}(2r)!\sum_{n\in\N\smallsetminus\{0\}}\frac{\cos(2n\pi t)}{(2n\pi)^{2r}},~~\forall t\in\R~~\textrm{and}~~\forall r\in\N\smallsetminus\{0\}.
  $$

Let $s\in]0,\frac12]$ be fixed for the rest of this subsection until further notice, the case $s=0$ being trivial. Let $\epsilon>0$ be also fixed.

Since the function $g_\epsilon(\cdot,s)$ is even on $\R$, $\frac{\pr^{2n-1}}{\pr t^{2n-1}}g_\epsilon(t,s)\big|_{t=0}=0$, for any $n\in\N\smallsetminus\{0\}$, and $g_\epsilon(0,s)=2\pi s$, the Euler-Maclaurin formula, for any $N\in\N$ and any $r\in\N\smallsetminus\{0\}$ writes
  \begin{equation}
\sum_{p=1}^Ng_\epsilon(p,s)=\int_{[0,N]}\!\!\dd t~\!g_\epsilon(t,s)+\frac{g_\epsilon(\cdot,s)(N)}2-\pi s+\sum_{n=1}^r\frac{b_{2n}}{(2n)!}\frac{\pr^{2n-1}}{\pr t^{2n-1}}g_\epsilon(t,s)\big|_{t=N}+R_r^{N,\epsilon}(s), \tag{$\bigstar_{\epsilon,r,N}$} \label{bigstar_epsilon,r,N}
  \end{equation}
where the remainder is given by
  $$
R_r^{N,\epsilon}(s)=-\int_{[0,N]}\!\!\dd t~\!\frac{\widetilde B_{2r}(t)}{(2r)!}\frac{\pr^{2r}}{\pr t^{2r}}g_\epsilon(t,s).
  $$
The mistake in \cite{Cheon_1988} is due to the fact that both limits $r,N\rightarrow\infty$ are taken without justifications for any $s\in\R\smallsetminus\{0\}$. For $s=0$, this limit yields a vanishing remainder. However, for $s\in\R\smallsetminus\{0\}$, care must be exercised, since the dominated convergent theorem does not actually apply in this case. As we shall see, it is easy to prove that it is possible to take the limit $N\rightarrow\infty$. But, the limit $R_r^\epsilon(s)$ of the remainder $R_r^{N,\epsilon}(s)$ does not go to zero in the limit $r\rightarrow\infty$, for any $s\in\R$, even after having taken the limit $\epsilon\rightarrow0^+$. \ \\ \\
  \begin{prop}~ \label{proposition_3}

The limit $N\rightarrow\infty$ of Eq. (\ref{bigstar_epsilon,r,N}) exists, does not depend on $r\in\N\smallsetminus\{0\}$, and is given by
    \begin{equation}
g_\epsilon(s)=\int_{\R_+}\!\!\dd t~\!g_\epsilon(t,s)-\pi s+R_1^\epsilon(s). \tag{$\bigstar_\epsilon$} \label{bigstar_epsilon}
    \end{equation}
    \begin{proof}
Let $r\in\N\smallsetminus\{0\}$.

First, $g_\epsilon(\cdot,s)\in\mathcal S(\R)$ implies that $g_\epsilon(\cdot,s)(N),\frac{\pr^{2n-1}}{\pr t^{2n-1}}g_\epsilon(t,s)\big|_{t=N}\xrightarrow[N\rightarrow\infty]{}0$, $\forall n\in\N\smallsetminus\{0\}$.

Second, $g_\epsilon(\cdot,s)\in\mathcal L^1(\Omega,\dd\mu)$, where the set $\Omega$ is either $\R_+$ or $\N\smallsetminus\{0\}$, fitted with its natural measure $\mu$, being respectively the Lebesgue measure or the so-called Dirac comb $\Delta=\sum_{p\in\Z}\delta_p$, $\delta_p$ being the Dirac measure at $p\in\Z$. Then, from dominated convergence theorem, $\int_{\Omega\cap[0,N]}\dd\mu~\!g_\epsilon(\cdot,s)\xrightarrow[N\rightarrow\infty]{}\int_\Omega\dd\mu~\!g_\epsilon(\cdot,s)$, so that
  $$
\sum_{p=1}^Ng_\epsilon(p,s)\xrightarrow[N\rightarrow\infty]{}\sum_{p\in\N\smallsetminus\{0\}}g_\epsilon(p,s)~~\textrm{and}~~\int_{[0,N]}\!\!\dd t~\!g_\epsilon(t,s)\xrightarrow[N\rightarrow\infty]{}\int_{\R_+}\!\!\dd t~\!g_\epsilon(t,s).
  $$

Third, $\lVert\widetilde B_{2r}\rVert_\infty^\R=|b_{2r}|<\infty$ and $g_\epsilon(\cdot,s)\in\mathcal S'(\R)$, hence $\widetilde B_{2r}\frac{\pr^{2r}}{\pr t^{2r}}g_\epsilon(\cdot,s)\in\mathcal L^1(\R,\dd t)$, and
  $$
R_r^{N,\epsilon}(s)\xrightarrow[N\rightarrow\infty]{}-\int_{\R_+}\!\!\dd t~\!\frac{\widetilde B_{2r}(t)}{(2r)!}\frac{\pr^{2r}}{\pr t^{2r}}g_\epsilon(t,s)=R_r^\epsilon(s).
  $$

Then, the limit $N\rightarrow\infty$ is well-defined, and yields, $\forall r\in\N\smallsetminus\{0\}$
  $$
g_\epsilon(s)=\int_{\R_+}\!\!\dd t~\!g_\epsilon(t,s)-\pi s+R_r^\epsilon(s)=\int_{\R_+}\!\!\dd t~\!g_\epsilon(t,s)-\pi s+R_1^\epsilon(s).
  $$
    \end{proof}
  \end{prop}

  \begin{prop}~ \label{proposition_4}

The limit $\epsilon\rightarrow0^+$ of Eq. (\ref{bigstar_epsilon}) exists, and
    \begin{equation}
g(s)=g_0(s),~~\forall s\in[0,\tfrac12]. \tag{$\bigstar$} \label{bigstar}
    \end{equation}
Moreover, there exists $C>0$, such that $\lVert g_\cdot(s)\rVert_\infty^{[0,1]}\leq C$, $\forall s\in[0,\tfrac12]$, and $g_\epsilon\xrightarrow[\epsilon\rightarrow0^+]{}g$ in $\mathcal S'(\R)$.
    \begin{proof}
First, we recall that it is proved in proposition {\bf\ref{proposition_1}} that the limit $\epsilon\rightarrow0^+$ of the quantities $g_\epsilon(s)$ and $\int_{\R_+}\!\!\dd t~\!g_\epsilon(t,s)$ exist and are  respectively $g_0(s)$ and $\frac\pi2\sign(s)$. Only the limit $\epsilon\rightarrow0^+$ of the remainder $R_1^\epsilon(s)$ is left to be justified.

Since the Fourier series representation of the function $\widetilde B_2$ are normally convergent on $\R$, we can write
  $$
R_1^\epsilon(s)=-\int_{\R_+}\!\!\dd t~\!\frac{\widetilde B_2(t)}{2}\frac{\pr^2}{\pr t^2}g_\epsilon(t,s)=-2\sum_{n\in\N\smallsetminus\{0\}}\int_{\R_+}\!\!\dd t~\!\frac{\cos2n\pi t}{(2n\pi)^2}\frac{\pr^2}{\pr t^2}g_\epsilon(t,s)=\sum_{n\in\N\smallsetminus\{0\}}h_\epsilon(n,s).
  $$
After two integrations by parts, for any $\epsilon>0$ and any $n\in\N\smallsetminus\{0\}$, since $s\in[0,\frac12]$, we get
  $$
h_\epsilon(n,s)=\frac\pi2\!\left\{\mathrm{erf}\!\left(\frac\pi{\sqrt\epsilon}(n+s)\right)-\mathrm{erf}\!\left(\frac\pi{\sqrt\epsilon}(n-s)\right)\right\}.
  $$
Then, for any $n\in\N\smallsetminus\{0\}$ and for any $\epsilon>0$, we have the following (in)equalities
  \begin{align*}
0\leq h_\epsilon(n,s)&=\sqrt\pi\int_{\frac\pi{\sqrt\epsilon}(n-s)}^{\frac\pi{\sqrt\epsilon}(n+s)}\!\!\dd t~\!\e^{-t^2}\leq\sqrt\frac{\pi^3}\epsilon
\int_{-\frac12}^{\frac12}\!\!\dd t~\!\e^{-\frac{\pi^2}\epsilon(t+n)^2}\leq\sqrt\frac{\pi^3}\epsilon\e^{-\frac{\pi^2}\epsilon n^2}\int_{-\frac12}^{\frac12}\!\!\dd t~\!\e^{-\frac{\pi^2}\epsilon t^2}\e^{-2\frac{\pi^2}\epsilon nt}
\\
&\leq\sqrt\frac{\pi^3}\epsilon\e^{-\frac{\pi^2}\epsilon n(n-1)}\int_\R\!\!\dd t~\!\e^{-\frac{\pi^2}\epsilon t^2}=\pi\e^{-\frac{\pi^2}\epsilon(n-1)}.
  \end{align*}
Then, we deduce that the function $\epsilon\longmapsto h_\epsilon(n,s)$ is continuous on $\R_+$ by continuous extension with $h_0(n,s)=0$, because $h_\cdot(n,s)$ is continuous on $\R\smallsetminus\{0\}_+$, and $0\leq h_\epsilon(n,s)\leq\pi\e^{-\frac{\pi^2}\epsilon(n-1)}\xrightarrow[\epsilon\rightarrow0^+]{}0$.

Second, we have $\lVert
h_\cdot(n,s)\rVert^{[0,1]}_\infty\leq\pi\e^{-\pi^2(n-1)}$, $\forall n\in\N\smallsetminus\{0\}$, implying that the series $\epsilon\longmapsto\sum h_\epsilon(n,s)$ are normally convergent on $[0,1]$, and $\lVert R_1^\cdot(s)\rVert_\infty^{[0,1]}\leq\pi\sum_{n\in\N}\e^{-\pi^2n}=\frac\pi{1-\e^{-\pi^2}}$. Finally, $\epsilon\longmapsto R_1^\epsilon(s)\in\mathcal C^0([0,1])$, and $R_1^\epsilon(s)\xrightarrow[\epsilon\rightarrow0^+]{}\sum_{n\in\N\smallsetminus\{0\}}h_0(n,s)=0$. The limit $\epsilon\rightarrow0^+$ of Eq. (\ref{bigstar_epsilon}) is therefore well-defined and given by
  $$
g_0(s)=\pi(\tfrac12-s)=g(s),~~\forall s\in]0,\tfrac12].
  $$
Since $g_0$ and $g:x\longmapsto\arctan\tan\pi(\tfrac12-s)\chi_{\R\smallsetminus\Z}(x)$ are both 1-periodic odd functions on $\R$ and coincide on $[0,\frac12]$, they are equal on $\R$.

We have also proved here the existence of the constant $C>0$ needed by proposition {\bf\ref{proposition_2}}. For any $\epsilon>0$, $\int_{\R_+}\!\!\dd t~\!g_\epsilon(t,s)=\frac\pi2\sign(s)\!~\mathrm{erf}\frac\pi{\sqrt\epsilon}|s|$ and $\int_{\R_+}\!\!\dd t~\!g_0(t,s)=\frac\pi2\sign(s)$, then $\lVert\int_{\R_+}\!\!\dd t~\!g_\cdot(t,s)\rVert_\infty^{[0,1]}\leq\frac\pi2$. So that, for any $s\in[0,\frac12]$,
  $$
\lVert g_\cdot(s)\rVert_\infty^{[0,1]}\leq\left\lVert\int_{\R_+}\!\!\dd t~\!g_\cdot(t,s)\right\rVert_\infty^{[0,1]}+\pi s+\lVert R_1^\cdot(s)\rVert_\infty^{[0,1]}\leq\frac{2\pi}{1-\e^{-\pi^2}}=C.
  $$

Finally, to sum up, we have just proved that $g_\epsilon\xrightarrow[\epsilon\rightarrow0^+]{}g$, point-wise in $\R$, and in $\mathcal S'(\R)$.
    \end{proof}
  \end{prop}
{\bf Remark~~}For  $s\in]0,\tfrac12]$, we have proved that the remainder $R^\epsilon_1(s)$ goes to zero in the limit $\epsilon\rightarrow0^+$. This becomes not true as soon as we consider $s\in\R\smallsetminus\{0\}$, but the same reasoning can be done to determine the new limit.

Using the same notations, we have that for any $n\in\N\smallsetminus\{0\}$, $\epsilon>0$ and $s\in\R\smallsetminus\{0\}$
  \begin{align*}
h_\epsilon(n,s)&=\frac\pi2\sign(s)\!\left\{\mathrm{erf}\!\left(\frac\pi{\sqrt\epsilon}(n+|s|)\right)-\sign(n-|s|)\mathrm{erf}\!\left(\frac\pi{\sqrt\epsilon}\big|n-|s|\big|\right)\right\}\!
\\
&\xrightarrow[\epsilon\rightarrow0^+]{}\frac\pi2\sign(s)\!\left\{1-\sign(n-|s|)\right\}=h_0(n,s).
  \end{align*}
Now, for any $n\geq|s|$, we have that $\lVert h_\cdot(n,s)\rVert^{[0,1]}_\infty\leq\pi\e^{-\pi^2(n-2|s|)}$. Reasoning as above, we conclude that $\epsilon\longmapsto R_1^\epsilon(s)\in\mathcal C^0([0,1])$, for any $s\in\R\smallsetminus\{0\}$, and
  $$
R_1^\epsilon(s)\xrightarrow[\epsilon\rightarrow0^+]{}\sum_{n\in\N\smallsetminus\{0\}}h_0(n,s)=\sum_{n=1}^{[|s|]}h_0(n,s)=\pi\left\{
    \begin{array}{ccl}
[s] & \textrm{if} & s\notin\Z
\\
s-\tfrac{\sign(s)}2 & \textrm{if} & s\in\Z
    \end{array}
  \right.\!\!\neq0,
  $$
where the function $[\cdot]$: $x\longmapsto[x]$ is the integer part function on $\R$. The limit $\epsilon\rightarrow0^+$ of Eq. (\ref{bigstar_epsilon}) is still well-defined, and yields
  $$
g_0(s)=\pi\left\{
    \begin{array}{ccl}
\frac{\sign(s)}2+[s]-s & \textrm{if} & s\notin\Z
\\
0 & \textrm{if} & s\in\Z
    \end{array}
  \right.\!\!.
  $$
Once again, $g$ and $g_0$ are equal on $\R$, because they are both odd and 1-periodic on $\R$, and coincide with the function $x\longmapsto\pi(\frac12-x)$ on $[0,\frac12]$.
\section{The mathematics of the Casimir effect} \label{appendix_B}
\subsection{Density of states in the presence of a Casimir device}
In this appendix, we present in details the general result on the density of states in vacuum or in the presence of a Casimir device. Let $F$ be a non-zero function and consider the following formal quantity 
  $$
H^F_\phi(L,d)=\sum_{(\nn_\parallel,n_\perp)\in\Z^3}F(k_{\nn_\parallel n_\perp})\phi\!\!\left(\frac{k_{\nn_\parallel n_\perp}}{\kappa_\phi}\right)\!=\sum_{(\nn_\parallel,n_\perp)\in\Z^3}\frac{f(k_{\nn_\parallel n_\perp})}{k_{\nn_\parallel n_\perp}}f_\gamma(k_{\nn_\parallel n_\perp})\phi\!\!\left(\frac{k_{\nn_\parallel n_\perp}}{\kappa_\phi}\right)\!, \label{H^F_phi(L,d)_sum}
  $$
where the two functions $f^\beta$ and $f_\gamma$ satisfy two assumptions:
  \begin{enumerate}[label={\bf\roman*.~}]
\item $f^\beta$: $s\longmapsto f(\kappa_d s)$ is a dimensioned function of class $\mathcal C^\infty$ on $\R_+$, of dimension $k^{-\beta}$, where $\beta\geq0$ is defined by the asymptotic behavior $f(k)=O(k^{-\beta})$ for $k\rightarrow\infty$.
\item $f_\gamma$: $s\longmapsto s^\gamma$, with $\gamma\in\R$, is a dimensionless function of class $\mathcal C^\infty$ at least on $\R\smallsetminus\{0\}$.
  \end{enumerate}
Note that $(\beta,\gamma)$ is uniquely determined. Moreover, $f^\beta\in\mathcal C^\infty(\R_+)$ and its successive derivatives are bounded on $\R_+$.
%
%
%
\subsection{Integral representation of $H^F_\phi(L,d)$}
Euler-Maclaurin or Poisson formulas \cite{Gourdon_Whittaker}, being the same equation once written in the tempered distributions formalism, cannot be used in the previous expression, because the function $s_\perp\longmapsto F\!\left(\kappa_ds_{\parallel\perp}\right)\phi\!\left(\frac{\kappa_d}{\kappa_\phi}s_{\parallel\perp}\right)$, is not differentiable at $s_\perp=0$, when $s_\parallel=0$, with $s_{\parallel\perp}=\sqrt{\ss^2_\parallel+s_\perp^2}$.

Let be $0<\delta\leq\eta$, and consider the regularized expression of $H^F_\phi(L,d)$
  $$
H^F_{\phi,\delta,\eta}(L,d)=\sum_{(\nn_\parallel,n_\perp)\in\Z^3}\frac{f\!\left(\sqrt{\kk^2_{\nn_\parallel n_\perp}\!+(\kappa_d\delta)^2}\right)}{\sqrt{\kk^2_{\nn_\parallel n_\perp}\!+(\kappa_d\delta)^2}}f_\gamma\!\left(\sqrt{\kk^2_{\nn_\parallel n_\perp}\!+(\kappa_d\eta)^2}\right)\phi\!\!\left(\frac{\sqrt{\kk^2_{\nn_\parallel n_\perp}\!+(\kappa_d\eta)^2}}{\kappa_\phi}\right)\!\!.
  $$
For $\ss_\parallel\in\R^2$ and $s_\perp\in\R$, denoting for convenience $s^{\parallel\perp}_\delta=\sqrt{\ss^2_\parallel+s_\perp^2+\delta^2}$, the regularized function $(\ss_\parallel,s_\perp)\longmapsto\frac{f^\beta(s^{\parallel\perp}_\delta)}{s^{\parallel\perp}_\delta}\Phi_\gamma(s^{\parallel\perp}_\eta)$ belongs to $\mathcal S(\R^2\times\R)$ --- the function $\Phi_\gamma$: $s\longmapsto f_\gamma(s)\phi(\frac{\kappa_d}{\kappa_\phi}s)$ is of class $\mathcal C^\infty$ and rapidly decreasing on $\R\smallsetminus[-\eta,\eta]$, for any $\eta>0$. So that, Euler-Maclaurin and Poisson formulas hold.
\ \\ \ \\
{\bf Remark~~}In the limit $L\gg d$, the sum over the quantum numbers $\nn_\parallel\in\Z^2$ in the previous expression may be replaced by an integral over $\R^2$ with the dimensionless measure $\frac{L^2}{(2\pi)^2}\dd^2\kk_\parallel$. The equality of these two quantities should be understood in the sense of asymptotic series. For simplicity, consider the function $G$: $k\longmapsto\frac{f\!\left(\sqrt{k^2+(\kappa_d\delta)^2}\right)}{\sqrt{k^2+(\kappa_d\delta)^2}}f_\gamma\!\left(\sqrt{k^2+(\kappa_d\eta)^2}\right)\phi\!\left(\frac{\sqrt{k^2+(\kappa_d\eta)^2}}{\kappa_\phi}\right)$, and call $G_1$ the function  $s\longmapsto\frac{f^\beta(s_\delta)}{s_\delta}\Phi_\gamma(s_\eta)$, with $s_{\pm\delta}=\sqrt{s^2\pm\delta^2}$. Since $G$ is even, the Euler-Maclaurin formula yields, for any $r\in\N\smallsetminus\{0\}$
  $$
\sum_{n\in\Z}G\!\left(\frac{2\pi}Ln\right)\!=\int_\R\dd t\!~G\!\left(\frac{2\pi}Lt\right)\!-\!\int_\R\dd t~\!\frac{\widetilde B_{2r}(t)}{(2r)!}\frac{\dd^{2r}}{\dd t^{2r}}G\!\left(\frac{2\pi}Lt\right)\!
  $$
Moreover, $\frac L{2\pi}\int_\R\dd k\!~G(k)=\kappa_d^\gamma\frac L{2d}\int_\R\dd s\!~G_1(s)$, we then focus on the quantity
  $$
\left(\kappa_d^\gamma\frac L{2d}\right)^{\!\!-1}\left|\!\int_\R\dd t~\!\frac{\widetilde B_{2r}(t)}{(2r)!}\frac{\dd^{2r}}{\dd t^{2r}}G\!\left(\frac{2\pi}Lt\right)\!\right|\leq2^{2r}\frac{|b_{2r}|}{(2r)!}\left(\frac dL\right)^{\!\!2r}\!C_{2r},
  $$
where the dimensioned constant $C_r=\int_\R\dd s\!~|G^{(r)}_1(s)|$ does not depend on $L$. This implies that for any $r\in\N\smallsetminus\{0\}$, the limit $\frac dL\rightarrow0$ can be taken, leading to
  $$
\sum_{n\in\Z}G\!\left(\frac{2\pi}Ln\right)\!=\kappa_d^\gamma\frac L{2d}\!\left\{\int_\R\dd s\!~G_1(s)+O\! \left[\!\left(\frac dL\right)^{\!\!2r}\!\right]\!\right\},
  $$
i.e. $\sum_{n\in\Z}G\!\left(\frac{2\pi}Ln\right)\!=\int_\R\dd k\!~G(k)$ in the sense of asymptotic series.

Here in particular, it is not possible to show that $2^{2r}\frac{|b_{2r}|}{(2r)!}\!\left(\frac dL\right)^{\!2r}C_{2r}\xrightarrow[r\rightarrow\infty]{}0$, for fixed $L\gg d$. In the neighborhood of $r\rightarrow\infty$, using the asymptotic behavior of the Bernoulli numbers $|b_{2r}|\sim 2\frac{(2r)!}{(2\pi)^{2r}}$, the asymptotic behavior of $2^{2r}\frac{|b_{2r}|}{(2r)!}\!\left(\frac dL\right)^{\!2r}C_{2r}$ is $2^{2r+1}\!\left(\frac d{2\pi L}\right)^{\!2r}C_{2r}$. However, the successive derivative in the constant $C_{2r}$ will generally produce factors of the order at least of $(2r)!$, which dominate the geometric dependence in $\frac d{2\pi L}$ in the remainder of Euler-Maclaurin formula for sufficiently large $r$.
\ \\
\par The Poisson formula is now applied to the function $s_\perp\longmapsto\frac{f^\beta(s^{\parallel\perp}_\delta)}{s^{\parallel\perp}_\delta}\Phi_\gamma(s^{\parallel\perp}_\eta)$, for any $\ss_\parallel\in\R^2$, and yields, after making the change of variables $\kk_\parallel\rightarrow\kappa_d\ss_\parallel$
  \begin{align}
H^F_{\phi,\delta,\eta}(L,d)&=L^2\frac{\kappa_d^{\gamma+1}}{4\pi^2}\sum_{n_\perp\in\Z}\int_{\R^2}\!\!\dd^2\ss_\parallel\frac{f^\beta\!\left(\sqrt{\ss^2_\parallel+n_\perp^2+\delta^2}\right)}{\sqrt{\ss_\parallel^2+n_\perp^2+\delta^2}}\Phi_\gamma\!\left(\frac{\kappa_d}{\kappa_\phi}\sqrt{\ss^2_\parallel+n_\perp^2+\eta^2}\right)\! \nt
\\
&=L^2\frac{\kappa_d^{\gamma+1}}{4\pi^2}\int_{\R^2}\!\!\dd^2\ss_\parallel\sum_{p\in\Z}\int_\R\dd s_\perp\frac{f^\beta(s^{\parallel\perp}_\delta)}{s^{\parallel\perp}_\delta}\Phi_\gamma(s^{\parallel\perp}_\eta)\e^{2\ii\pi ps_\perp}. \tag{$\bigstar^F_{\phi,\delta,\eta}$} \label{H^F_phi,delta,eta(L,d)}
  \end{align} \ \\
  \begin{lem}~ \label{lemma_1}

The function $(\ss_\parallel,n_\perp)\longmapsto\frac{f^\beta\!\left(\sqrt{\ss^2_\parallel+n_\perp^2+\delta^2}\right)}{\sqrt{\ss_\parallel^2+n_\perp^2+\delta^2}}\Phi_\gamma\!\left(\sqrt{\ss^2_\parallel+n_\perp^2+\eta^2}\right)$ is integrable on $(\R^2\times\Z,\dd^2\ss_\parallel\otimes\dd\Delta(n_\perp))$. The sum $\sum_{n_\perp\in\Z}$ and the integral $\int_{\R^2}\dd^2\ss_\parallel$ in Eq. (\ref{H^F_phi,delta,eta(L,d)}) can be inverted.
    \begin{proof}
For any $n_\perp\in\Z$, we can write, after performing the change of variables $s=\sqrt{s^2_\parallel+n_\perp^2+\eta^2}$
      \begin{align*}
\int_{\R^2}\!\!\dd^2\ss_\parallel \left|\frac{f^\beta\!\left(\sqrt{\ss^2_\parallel+n_\perp^2+\delta^2}\right)}{\sqrt{\ss_\parallel^2+n_\perp^2+\delta^2}}\Phi_\gamma\!\left(\sqrt{\ss^2_\parallel+n_\perp^2+\eta^2}\right)\right|\!&\leq2\pi\lVert f^\beta\rVert_\infty^{\R_+}\int_{\R_+}\!\!\frac{\dd s_\parallel\!~s_\parallel}{\sqrt{s_\parallel^2+n_\perp^2}}\!~\left|\Phi_\gamma\!\left(\sqrt{s^2_\parallel+n_\perp^2+\eta^2}\right)\right|
\\
&\leq2\pi\lVert f^\beta\rVert_\infty^{\R_+}\int_{\sqrt{n_\perp^2+\eta^2}}^\infty\!\frac{\dd s\!~s}{s_{-\eta}}\left|\Phi_\gamma(s)\right|
\\
&\leq2\pi\lVert f^\beta\rVert_\infty^{\R_+}\int_{\sqrt{n_\perp^2+\eta^2}}^\infty\dd s\!~\sqrt{\frac s{s-\eta}}\left|\Phi_\gamma(s)\right|
\\
&\leq2\pi\lVert f^\beta\rVert_\infty^{\R_+}\int_{|n_\perp|}^\infty\dd s\left|\Phi_\gamma(s)\right|.
      \end{align*}
The last inequality is obtained by integration by parts
      \begin{align*}
\int_{\sqrt{n_\perp^2+\eta^2}}^\infty\dd s\!~\sqrt{\frac s{s-\eta}}\left|\Phi_\gamma(s)\right|\!&=2\sqrt{\sqrt{n_\perp^2+\eta^2}-\eta}\int_{\sqrt{n_\perp^2+\eta^2}}^\infty\dd t\!~\sqrt t\left|\Phi_\gamma(t)\right|
\\
&~~~~~~+2\int_{\sqrt{n_\perp^2+\eta^2}}^\infty\dd s\!~\sqrt{s-\eta}\int_s^\infty\dd t\!~\sqrt t\left|\Phi_\gamma(t)\right|
\\
&\leq2\sqrt{|n_\perp|}\int_{|n_\perp|}^\infty\dd t\!~\sqrt t\left|\Phi_\gamma(t)\right|+2\int_{|n_\perp|}^\infty\dd s\!~\sqrt s\int_s^\infty\dd t\!~\sqrt t\left|\Phi_\gamma(t)\right|
\\
&=\int_{|n_\perp|}^\infty\dd s\left|\Phi_\gamma(s)\right|.
      \end{align*}     
      
Furthermore, $s^3\left|\Phi_\gamma(s)\right|\xrightarrow[s\rightarrow\infty]{}0$, then $\exists N_\perp\in\N\smallsetminus\{0\}$, such that $\left|\Phi_\gamma(s)\right|\leq\frac1{s^3}$, $\forall s\geq N_\perp$. Then, for any $n_\perp\in\Z$ such that $|n_\perp|\geq N_\perp$, we have
  $$
\int_{\R^2}\!\!\dd^2\ss_\parallel\left|\frac{f^\beta\!\left(\sqrt{\ss^2_\parallel+n_\perp^2+\delta^2}\right)}{\sqrt{\ss_\parallel^2+n_\perp^2+\delta^2}}\Phi_\gamma\!\left(\sqrt{\ss^2_\parallel+n_\perp^2+\eta^2}\right)\right|\!\leq\frac{2\pi}3\frac{\lVert f^\beta\rVert_\infty^{\R_+}}{n^2_\perp},
  $$
which shows that the series $\sum\int_{\R^2}\dd^2\ss_\parallel\left|\frac{f^\beta\!\left(\sqrt{\ss^2_\parallel+n_\perp^2+\delta^2}\right)}{\sqrt{\ss_\parallel^2+n_\perp^2+\delta^2}}\Phi_\gamma\!\left(\sqrt{\ss^2_\parallel+n_\perp^2+\eta^2}\right)\right|\!$ are convergent, and by Fubini-Tonelli theorem
      \begin{align*}
&\int_{\R^2\times\Z^*}\dd^2\ss_\parallel\otimes\dd\Delta(n_\perp)\left|\frac{f^\beta\!\left(\sqrt{\ss^2_\parallel+n_\perp^2+\delta^2}\right)}{\sqrt{\ss_\parallel^2+n_\perp^2+\delta^2}}\Phi_\gamma\!\left(\sqrt{\ss^2_\parallel+n_\perp^2+\eta^2}\right)\right|\!
\\
=&\sum_{n_\perp\in\Z^*}\int_{\R^2}\!\!\dd^2\ss_\parallel\left|\frac{f^\beta\!\left(\sqrt{\ss^2_\parallel+n_\perp^2+\delta^2}\right)}{\sqrt{\ss_\parallel^2+n_\perp^2+\delta^2}}\Phi_\gamma\!\left(\sqrt{\ss^2_\parallel+n_\perp^2+\eta^2}\right)\right|\!<\infty,
      \end{align*}
{\it i.e.} $(\ss_\parallel,n_\perp)\longmapsto\frac{f^\beta\!\left(\sqrt{\ss^2_\parallel+n_\perp^2+\delta^2}\right)}{\sqrt{\ss_\parallel^2+n_\perp^2+\delta^2}}\Phi_\gamma\!\left(\sqrt{\ss^2_\parallel+n_\perp^2+\eta^2}\right)\!\in\mathcal L^1(\R^2\times\Z,\dd^2\ss_\parallel\otimes\dd\Delta(n_\perp))$, where $\dd^2\ss_\parallel\otimes\dd\Delta(n_\perp)$ is the usual product mesure on $\R^2\times\Z$. This leads to the inversion result by Fubini theorem.
    \end{proof}
  \end{lem} \ \\
\noindent\cor The function $\delta\rightarrow H^F_{\phi,\delta,\eta}(L,d)$ is continuous on $[0,\eta]$.
  \begin{proof}
The function $\delta\rightarrow\frac{f^\beta\!\left(\sqrt{\ss^2_\parallel+n_\perp^2+\delta^2}\right)}{\sqrt{\ss_\parallel^2+n_\perp^2+\delta^2}}\Phi_\gamma\!\left(\sqrt{\ss^2_\parallel+n_\perp^2+\eta^2}\right)$ is continuous on $[0,\eta]$. The result is deduced directly from the proof of lemma {\bf\ref{lemma_1}}, which suggests that for any $\ss_\parallel\in\R^2$,
  $$
\left|\frac{f^\beta\!\left(\sqrt{\ss^2_\parallel+n_\perp^2+\delta^2}\right)}{\sqrt{\ss_\parallel^2+n_\perp^2+\delta^2}}\Phi_\gamma\!\left(\sqrt{\ss^2_\parallel+n_\perp^2+\eta^2}\right)\right|\leq\frac{\lVert f^\beta\rVert_\infty^{\R_+}}{\sqrt{\ss_\parallel^2+N_\perp^2}(\ss_\parallel^2+n_\perp^2+\eta^2)^2}\leq\frac{\lVert f^\beta\rVert_\infty^{\R_+}}{n_\perp^2(\ss_\parallel^2+N_\perp^2)^{\frac32}},~~\forall|n_\perp|\geq N_\perp;
  $$
where $(\ss_\parallel,n_\perp)\longmapsto n_\perp^{-2}(\ss_\parallel^2+N_\perp^2)^{-\frac32}\chi_{\R\smallsetminus]-N_\perp,N_\perp[}(n_\perp)\in\mathcal L^1(\R^2\times\Z^*,\dd\ss_\parallel\otimes\dd\Delta(n_\perp))$, and $(\ss_\parallel,n_\perp)\longmapsto(\ss_\parallel^2+\eta^2)^{-\frac12}\left|\Phi_\gamma\!\left(\sqrt{\ss^2_\parallel+n_\perp^2+\eta^2}\right)\right|\chi_{[-N_\perp,N_\perp]}(n_\perp)\in\mathcal L^1(\R^2\times\Z^*,\dd\ss_\parallel\otimes\dd\Delta(n_\perp))$.
  \end{proof} \ \\

Let now $\epsilon>0$, and consider the fully regularized expression
  \begin{equation}
H^F_{\phi,\delta,\eta,\epsilon}(L,d)=L^2\frac{\kappa_d^{\gamma+1}}{4\pi^2}\int_{\R^2}\!\!\dd^2\ss_\parallel\sum_{p\in\Z}\int_\R\dd s_\perp\frac{f^\beta(s^{\parallel\perp}_\delta)}{s^{\parallel\perp}_\delta}\Phi_\gamma(s^{\parallel\perp}_\eta)\e^{2\ii\pi ps_\perp}\e^{-\epsilon p^2} \tag{$\bigstar^F_{\phi,\delta,\eta,\epsilon}$} \label{H^F_phi,delta,eta,epsilon(L,d)}
  \end{equation}
As we will see later, the limits $\delta,\epsilon\rightarrow0^+$ in Eq. (\ref{H^F_phi,delta,eta,epsilon(L,d)}) are well-defined, whereas it will be not possible to take the limit $\eta\rightarrow0^+$ in Eq. (\ref{H^F_phi,delta,eta,epsilon(L,d)}) and then Eq. (\ref{H^F_phi,delta,eta(L,d)}), when the function $\Phi_\gamma$ will not satisfy the assumptions of proposition {\bf\ref{proposition_6}}. \ \\ \\

  \begin{lem}~ \label{lemma_2}

The function $(\ss_\parallel,s_\perp,p)\longmapsto\frac{f^\beta(s^{\parallel\perp}_\delta)}{s^{\parallel\perp}_\delta}\Phi_\gamma(s^{\parallel\perp}_\eta)\e^{2\ii\pi ps_\perp}\e^{-\epsilon p^2}$ is integrable on $(\R^2\times\R\times\Z,\dd^2\ss_\parallel\otimes\dd s_\perp\otimes\dd\Delta(p))$. And, the sum $\sum_{n_\perp\in\Z}$ and the integrals $\int_{\R^2}\dd^2\ss_\parallel$ and $\int_\R\dd s_\perp$ in Eq. (\ref{H^F_phi,delta,eta,epsilon(L,d)}) can be inverted.
    \begin{proof}
For any $\ss_\parallel\in\R^2$, $s_\perp\in\R$ and $p\in\Z$, we have
  $$
\left|\frac{f^\beta(s^{\parallel\perp}_\delta)}{s^{\parallel\perp}_\delta}\Phi_\gamma(s^{\parallel\perp}_\eta)\e^{2\ii\pi ps_\perp}\e^{-\epsilon p^2}\right|\leq\lVert f^\beta\rVert_\infty^{\R_+}\frac{\left|\Phi_\gamma(s^{\parallel\perp}_\eta)\right|}{s^{\parallel\perp}}\e^{-\epsilon|p|}.
  $$
where, by Fubini-Tonelli theorem, and after performing the change of variable $s\rightarrow s_\eta$,
  $$
\int_{\R^2\times\R\times\Z}\dd^2\ss_\parallel\otimes\dd s_\perp\otimes\dd\Delta(p)\left|\Phi_\gamma(s^{\parallel\perp}_\eta)\right|\e^{-\epsilon |p|}\leq\frac2{1-\e^{-\epsilon}}\int_{\R^2\times\R}\dd^2\ss_\parallel\otimes\dd s_\perp\frac{\left|\Phi_\gamma(s^{\parallel\perp}_\eta)\right|}{s^{\parallel\perp}}\leq\frac{8\pi C_\eta}{1-\e^{-\epsilon}},
  $$
with $C_\eta=\int_{\R_+}\!\!\dd s\!~s|\Phi_\gamma(s_\eta)|<\infty$. Then, $(\ss_\parallel,s_\perp,p)\longmapsto\frac{f^\beta(s^{\parallel\perp}_\delta)}{s^{\parallel\perp}_\delta}\Phi_\gamma(s^{\parallel\perp}_\eta)\e^{2\ii\pi ps_\perp}\e^{-\epsilon p^2}\in\mathcal L^1(\R^2\times\R\times\Z,\dd^2\ss_\parallel\otimes\dd s_\perp\otimes\dd\Delta(p))$, and the inversion property is obtained by Fubini theorem.
    \end{proof}
  \end{lem} \ \\
  \begin{lem}~ \label{lemma_3}

The family $(H_\epsilon)_\epsilon$, where $H_\epsilon$: $s\longmapsto\sum_{p\in\Z}\e^{2\ii\pi ps}\e^{-\epsilon p^2}$, for any $\epsilon>0$, is convergent in $\mathcal S'(\R)$, and goes to the Dirac comb $\Delta=\widehat\Delta$.
    \begin{proof}
    
Let the functions $h_\epsilon$: $s\longmapsto\frac1{\sqrt\epsilon}h(\frac s{\sqrt\epsilon})$ and $h$: $s\longmapsto\sqrt\pi\e^{-(\pi s)^2}$, which satisfies $\int_\R\dd s\!~h(s)=1$, such that $h_\epsilon\xrightarrow[\epsilon\rightarrow0^+]{}\delta_0$ in $\mathcal S'(\R)$.
\ \\ \ \\
{\bf Remark~~}{\rm It is well-known  that the Dirac comb $\Delta$ is equal to its Fourier transform $\widehat\Delta(s)=\sum_{p\in\Z}\e^{2\ii\pi ps}$ in $\mathcal S'(\R)$.

Since $h\in\mathcal S(\R)$ and $h_\epsilon(s)=\int_\R\dd t\!~\e^{2\ii\pi st}\e^{-\epsilon t^2}$, inversion theorem holds and yields $\e^{-\epsilon s^2}=\int_\R\dd t\!~\e^{2\ii\pi st}h_\epsilon(t)$. Moreover, for any $\phi\in\mathcal S(\R)$, by the dominated convergence theorem,
  $$
\int_\R\dd s\!~h_\epsilon(s)\phi(s)=\int_\R\dd s\!~h(s)\phi(\sqrt\epsilon s)\xrightarrow[\epsilon\rightarrow0^+]{}\phi(0)=\int_\R\dd s\!~\delta_0(s)\phi(s)~~\Leftrightarrow~~h_\epsilon\xrightarrow[\epsilon\rightarrow0^+]{}\delta_0~~\textrm{in}~~\mathcal S'(\R).
  $$
From this result and since $(s,t)\longmapsto h_\epsilon(t)\phi(s)\in\mathcal L^1(\R^2,\dd s\dd t)$ we get
  $$
\int_\R\dd s\!~\phi(s)\xleftarrow[\epsilon\rightarrow0^+]{}\int_\R\dd s\!~\e^{-\epsilon s^2}\phi(s)=\int_\R\dd t\!~h_\epsilon(t)\widehat\phi(t)\xrightarrow[\epsilon\rightarrow0^+]{}\int_\R\dd t\!~\delta_0(t)\widehat\phi(t)~~\Leftrightarrow~~\widehat{\delta_0}=1~~\textrm{in}~~\mathcal S'(\R).
  $$
Using the translation property of the Fourier transform, we have $\widehat{\delta_p}(s)=\e^{2\ii\pi ps}$, for any $p\in\Z$, and by linearity of the Fourier transform in $\mathcal S'(\R)$, we obtain that $\widehat\Delta(s)=\sum_{p\in\Z}\e^{2\ii\pi ps}$. Finally, using Poisson formula for $\phi\in\mathcal S(\R)$, we deduce
  $$
\int_\R\dd s\!~\Delta(s)\phi(s)=\sum_{n\in\Z}\phi(n)=\sum_{p\in\Z}\widehat\phi(p)=\int_\R\dd s\!~\widehat\Delta(s)\phi(s)~~\Leftrightarrow~~\Delta=\widehat\Delta~~\textrm{in}~~\mathcal S'(\R).
  $$}
  
We now apply the same reasoning as in the proofs of propositions {\bf\ref{proposition_1}-\ref{proposition_4}} to the series $H_\epsilon$: $s\longmapsto\sum_{p\in\Z}\e^{2\ii\pi ps}\e^{-\epsilon p^2}$. Using the Euler-Maclaurin formula, we have, for any $s\in\R$,
  $$
H_\epsilon(s)=\int_\R\dd t\!~\e^{2\ii\pi st}\e^{-\epsilon t^2}+2\sum_{n\in\N\smallsetminus\{0\}}\int_\R\dd t\!~\cos2n\pi t\!~\e^{2\ii\pi st}\e^{-\epsilon t^2}=\sum_{n\in\Z}h_\epsilon(n-s)\geq0.
  $$
Moreover, $h_\epsilon$ satisfies $H_\epsilon(s)\leq\frac{\pi^2}3\left[\sqrt{\frac\pi\epsilon}\left(s^2+\frac1\epsilon+1\right)\!+|s|\right]$, $\forall s\in\R$. Then, for any $\phi\in\mathcal S(\R)$, $s\longmapsto H_\epsilon\phi\in\mathcal L^1(\R,\dd s)$, {\it i.e.} $H_\epsilon\in\mathcal S'(\R)$. And, by Fubini-Tonelli theorem,
  $$
\int_{\R\times\Z}\dd s\otimes\dd\Delta(n)\!~h_\epsilon(n-s)|\phi(s)|\!~=\int_\R\dd s\!~H_\epsilon(s)|\phi(s)|<\infty,
  $$
we obtain that $(s,n)\longmapsto h_\epsilon(n-s)\phi(s)\in\mathcal L^1(\R\times\Z,\dd s\otimes\dd\Delta(n))$. Then, by Fubini theorem,
  $$
\int_\R\dd s\!~H_\epsilon(s)\phi(s)=\int_{\R\times\Z}\!\!\dd s\otimes\dd\Delta(n)\!~h_\epsilon(n-s)\phi(s)=\int_{\R\times\Z}\!\!\dd s\otimes\dd\Delta(n)\!~h_\epsilon(s)\phi(s-n)=\int_\R h_\epsilon(s)\sum_{n\in\Z}\phi(s-n).
  $$
Since $s\longmapsto\sum_{n\in\Z}\phi(s-n)\in\mathcal S(\R)$, the limit $\epsilon\rightarrow0^+$ of the right-hand-side of this expression exists and we have
  $$
\int_\R\dd s\!~H_\epsilon(s)\phi(s)\xrightarrow[\epsilon\rightarrow0^+]{}\sum_{n\in\Z}\phi(n)=\int_\R\dd s\!~\Delta(s)\phi(s)~~\Leftrightarrow~~H_\epsilon\xrightarrow[\epsilon\rightarrow0^+]{}\Delta~~\textrm{in}~~\mathcal S'(\R).
  $$
    \end{proof}
  \end{lem} \ \\
  \begin{prop}~ \label{proposition_5}

The limit $\delta,\epsilon\rightarrow0^+$ of Eq. (\ref{H^F_phi,delta,eta,epsilon(L,d)}) exists
  $$
H^F_{\phi,\delta,\eta,\epsilon}(L,d)\xrightarrow[\delta,\epsilon\rightarrow0^+]{}H^F_{\phi,0,\eta,0}(L,d)=H^F_{\phi,\eta}(L,d).
  $$
    \begin{proof}
By lemma {\bf\ref{lemma_2}}, we have $(s_\perp,p)\longmapsto \frac{f^\beta(s^{\parallel\perp}_\delta)}{s^{\parallel\perp}_\delta}\Phi_\gamma(s^{\parallel\perp}_\eta)\e^{2\ii\pi ps_\perp}\e^{-\epsilon p^2}\in\mathcal L^1(\R\times\Z,\dd s_\perp\otimes\dd\Delta(p))$, for any $\epsilon>0$ and $\ss_\parallel\in\R^2$, then it is possible to inverse the sum $\sum_{p\in\Z}$ and the integral $\int_\R\dd s_\perp$
  $$
H^F_{\phi,\delta,\eta,\epsilon}(L,d)=L^2\frac{\kappa_d^{\gamma+1}}{4\pi^2}\int_{\R^2}\!\!\dd^2\ss_\parallel\int_\R\dd s_\perp\frac{f^\beta(s^{\parallel\perp}_\delta)}{s^{\parallel\perp}_\delta}H_\epsilon(s_\perp)\Phi_\gamma(s^{\parallel\perp}_\eta).
  $$
Moreover, by lemma {\bf\ref{lemma_3}}, we have, for almost any $(\ss_\parallel,s_\perp)\in\R^2\times\R$,
  $$
\left|\frac{f^\beta(s^{\parallel\perp}_\delta)}{s^{\parallel\perp}_\delta}H_\epsilon(s_\perp)\Phi_\gamma(s^{\parallel\perp}_\eta)\right|\leq\frac{\pi^2}3\lVert f^\beta\rVert_\infty^{\R_+}\frac{\Phi(s^{\parallel\perp}_\eta)}{s^{\parallel\perp}},
  $$
where $\Phi$: $s\longmapsto s\!\left[\sqrt{\frac\pi\epsilon}\left(s^2+\frac1\epsilon+1\right)\!+|s|\right]\!\left|\Phi_\gamma(s)\right|\in\mathcal L^1(\R,\dd s)$. This upper-bound almost surely on $\R^2\times\R$ is independent from $\delta\in]0,\eta]$ and is also valid for $\delta=0$. Then, $(\ss_\parallel,s_\perp)\longmapsto\frac{f^\beta(s^{\parallel\perp}_\delta)}{s^{\parallel\perp}_\delta}H_\epsilon(s_\perp)\Phi_\gamma(s^{\parallel\perp}_\eta)\in\mathcal L^1(\R^2\times\R,\dd^2\ss_\parallel\otimes\dd s_\perp)$, so that Fubini theorem yields, for any $\epsilon>0$ and any $\delta\in[0,\eta]$
  $$
H^F_{\phi,\delta,\eta,\epsilon}(L,d)=L^2\frac{\kappa_d^{\gamma+1}}{4\pi^2}\int_{\R^2\times\R}\!\!\dd^2\ss_\parallel\otimes\dd s_\perp\frac{f^\beta(s^{\parallel\perp}_\delta)}{s^{\parallel\perp}_\delta}H_\epsilon(s_\perp)\Phi_\gamma(s^{\parallel\perp}_\eta),
  $$
and the function $\delta\longmapsto H^F_{\phi,\delta,\eta,\epsilon}(L,d)$ is continuous $[0,\eta]$. Then, the limit $\delta\rightarrow0^+$ exists
  $$
H^F_{\phi,\delta,\eta,\epsilon}(L,d)\xrightarrow[\delta\rightarrow0^+]{}H^F_{\phi,0,\eta,\epsilon}(L,d).
  $$

But, for any $\delta\in[0,\eta]$, since $f^\beta\in\mathcal C^0(\R)$ is bounded on $\R$, $(\ss_\parallel,s_\perp)\longmapsto\frac{f^\beta(s^{\parallel\perp}_\delta)}{s^{\parallel\perp}_\delta}H_\epsilon(s_\perp)\in\mathcal S'(\R^2\times\R)$. Then, by lemma {\bf\ref{lemma_3}}, $\frac{f^\beta(s^{\parallel\perp}_\delta)}{s^{\parallel\perp}_\delta}H_\epsilon(s_\perp)\xrightarrow[\epsilon\rightarrow0^+]{}f^\beta(s^{\parallel\perp}_\delta)\Delta(s_\perp)$ in $S'(\R^2\times\R)$. Therefore, since $(\ss_\parallel,s_\perp)\longmapsto\Phi_\gamma(s^{\parallel\perp}_\eta)\in\mathcal S(\R^2,\R)$, $H^F_{\phi,\delta,\eta,\epsilon}(L,d)\xrightarrow[\epsilon\rightarrow0^+]{}H^F_{\phi,\delta,\eta,0}(L,d)$, where by properties of the Dirac comb $\Delta$, we have
      \begin{align*}
H^F_{\phi,\delta,\eta,0}(L,d)&=L^2\frac{\kappa_d^{\gamma+1}}{4\pi^2}\int_{\R^2\times\R}\!\!\dd^2\ss_\parallel\otimes\dd s_\perp\frac{f^\beta(s^{\parallel\perp}_\delta)}{s^{\parallel\perp}_\delta}\Delta(s_\perp)\Phi_\gamma(s^{\parallel\perp}_\eta).
\\
&=L^2\frac{\kappa_d^{\gamma+1}}{4\pi^2}\sum_{n_\perp\in\Z}\int_{\R^2}\!\!\dd^2\ss_\parallel\frac{f^\beta\!\left(\sqrt{\ss^2_\parallel+n_\perp^2+\delta^2}\right)}{\sqrt{\ss_\parallel^2+n_\perp^2+\delta^2}}\Phi_\gamma\!\left(\sqrt{\ss^2_\parallel+n_\perp^2+\eta^2}\right)\!=H^F_{\phi,\delta,\eta}(L,d).
      \end{align*}
This implies in particular that $H^F_{\phi,0,\eta,\epsilon}(L,d)\xrightarrow[\epsilon\rightarrow0^+]{}H^F_{\phi,0,\eta,0}(L,d)=H^F_{\phi,0,\eta}(L,d)$, the reciprocal being directly given by the corollary of lemma {\bf\ref{lemma_1}}.
    \end{proof}
  \end{prop} \ \\
  \begin{theo} For any $\delta\in[0,\eta]$, \label{theorem_1}
      \begin{align}
H^F_{\phi,\delta,\eta}(L,d)&=V\frac{\kappa_d^{\gamma+2}}{\pi^2}\int_{\R_+}\!\!\dd s\!~s^2\!\left[1+\frac{g(s)}{\pi s}\right]\frac{f^\beta(s_\delta)}{s_\delta}f_\gamma(s_\eta)\phi\!\left(\frac{\kappa_d}{\kappa_\phi}s_\eta\right) \nt
\\
&=V\int_{\R_+}\!\!\dd k\left[\rho(k)+\rho_d(k)\right]\frac{f(\sqrt{k^2+(\kappa_d\delta)^2})}{\sqrt{k^2+(\kappa_d\delta)^2}}\!f_\gamma\!\left(\sqrt{k^2+(\kappa_d\eta)^2}\right)\!\phi\!\left(\frac{\sqrt{k^2+(\kappa_d\eta)^2}}{\kappa_\phi}\right)\!. \tag{$\bigstar^F_{\phi,\delta,\eta}$} \label{H^F_phi,delta,eta}
      \end{align}
    \begin{proof}
By lemma {\bf\ref{lemma_2}}, the sum $\sum_{p\in\Z}$ and the integral $\int_\R\dd s_\perp$ can be inverted in Eq. (\ref{H^F_phi,delta,eta,epsilon(L,d)}), yielding
      \begin{align*}
H^F_{\phi,\delta,\eta,\epsilon}(L,d)&=L^2\frac{\kappa_d^{\gamma+1}}\pi\int_{\R_+}\!\!\dd s\!~s^2\!~\frac{f^\beta(s_\delta)}{s_\delta}\Phi_\gamma(s_\eta)\sum_{p\in\Z}\frac{\sin2\pi ps}{2\pi ps}\e^{-\epsilon p^2}
\\
&=L^2\frac{\kappa_d^{\gamma+1}}\pi\int_{\R_+}\!\!\dd s\!~s^2\!\left[1+\frac{g_\epsilon(s)}{\pi s}\right]\frac{f^\beta(s_\delta)}{s_\delta}\Phi_\gamma(s_\eta).
      \end{align*}
By the same reasoning as in the proof of proposition {\bf\ref{proposition_5}}, since by proposition {\bf\ref{proposition_4}}, $g_\epsilon\xrightarrow[\epsilon\rightarrow0^+]{}g$ in $\mathcal S'(\R)$, {\it i.e.} the limit $\epsilon\rightarrow0^+$ of the right-hand-side of the previous expression exists
  $$
H^F_{\phi,\delta,\eta}(L,d)\xleftarrow[\epsilon\rightarrow0^+]{}H^F_{\phi,\delta,\eta,\epsilon}(L,d)\xrightarrow[\epsilon\rightarrow0^+]{}L^2\frac{\kappa_d^{\gamma+1}}\pi\int_{\R_+}\!\!\dd s\!~s^2\!\left[1+\frac{g(s)}{\pi s}\right]\frac{f^\beta(s_\delta)}{s_\delta}\Phi_\gamma(s_\eta).
  $$
    \end{proof}
  \end{theo} \ \\
  \begin{prop}~ \label{proposition_6}

Assuming that $\gamma>-1$, then
    \begin{align}
H^F_\phi(L,d)&=V\frac{\kappa_d^{\gamma+2}}{\pi^2}\int_{\R_+}\!\!\dd s\!~s\!\left[1+\frac{g(s)}{\pi s}\right]\!f^\beta(s)f_\gamma(s)\phi\!\left(\frac{\kappa_d}{\kappa_\phi}s\right) \nt
\\
&=V\int_{\R_+}\!\!\dd k\left[\rho(k)+\rho_d(k)\right]F(k)\phi\!\left(\frac k{\kappa_\phi}\right)\!. \tag{$\bigstar^F_\phi$} \label{H^F_phi}
    \end{align}
    \begin{proof}
There exists $S>0$ such that $|\Phi_{\gamma+i}(s)|\leq\frac1{s^2}$, $\forall s\geq S$. Let $\mathbb K\subseteq\R_+$ a compact set, for any $\eta\in\mathbb K$ and $s\geq S$
  $$
\left|s^2\!\left[1+\frac{g(s)}{\pi s}\right]\!\frac{f^\beta(s_\eta)}{s_\eta}\Phi_\gamma(s_\eta)\right|\leq\lVert f^\beta\rVert_\infty^{\R_+}\left[|\Phi_{\gamma+1}(s_\eta)|+\frac{|\Phi_\gamma(s_\eta)|}2\right]\!\leq\frac32\frac{\lVert f^\beta\rVert_\infty^{\R_+}}{s^2}.
  $$
Moreover, $(\eta,s)\longmapsto\Phi_{\gamma+1}(s_\eta)\!\in\mathcal C^0(\mathbb K\times[0,S])$, and $f_\gamma\in\mathcal L^1([0,S],\dd s)$. Then, there exists $C_{\mathbb K,S}\geq0$ and $C'_{\mathbb K,S}\geq0$, such that, for any $\eta\in\mathbb K$ and any $s\in[0,S]$
    \begin{align*}
\left|s^2\!\left[1+\frac{g(s)}{\pi s}\right]\!\frac{f^\beta(s_\eta)}{s_\eta}\Phi_\gamma(s_\eta)\right|&\leq\lVert f^\beta\rVert_\infty^{\R_+}\left[|\Phi_{\gamma+1}(s_\eta)|+\frac{|f_\gamma(s_\eta)|}2\left|\phi\!\left(\frac{\kappa_d}{\kappa_\phi}s\right)\right|\right]
\\
&\leq\lVert f^\beta\rVert_\infty^{\R_+}\left[C_{\mathbb K,S}+\frac{C'_{\mathbb K,S}}2\left\{f_\gamma(s)+f_\gamma(S_{\max\mathbb K})\right\}\right]\!.
    \end{align*}
Then, $\eta\longmapsto H^F_{\phi,\eta,\eta}\in\mathcal C^0(\R_+)$, and the limit of Eq. ($\bigstar^F_{\phi,\eta,\eta}$) is well-defined
  $$
H^F_{\phi,\eta,\eta}(L,d)\xrightarrow[\eta\rightarrow0^+]{}H^F_{\phi,0,0}(L,d)=H^F_\phi(L,d).
  $$
    \end{proof}
  \end{prop}

Let us identify the modification of the quantity $H^F_{\phi,\eta}(L,d)$ due to the Casimir plates as
    \begin{align}
H^{F,\mathrm{Casimir}}_{\phi,\eta}(L,d)&=V\frac{\kappa_d^{\gamma+2}}{\pi^3}\int_{\R_+}\!\!\dd s\!~g(s)f^\beta(s)f_\gamma(s_\eta)\phi\!\left(\frac{\kappa_d}{\kappa_\phi}s_\eta\right) \nt
\\
&=V\int_{\R_+}\!\!\frac{\dd k}k\!~\rho_d(k)f(k)f_\gamma\!\left(\sqrt{k^2+(\kappa_d\eta)^2}\right)\!\phi\!\left(\frac{\sqrt{k^2+(\kappa_d\eta)^2}}{\kappa_\phi}\right)\!\!. \tag{$\bigstar^F_{\phi,\eta}$} \label{H^F_phi,eta}
    \end{align}
Theorem {\bf\ref{theorem_1}} and proposition {\bf\ref{proposition_6}} justify rigourously the heuristic reasoning made in subsection {\bf\ref{subsec_2_2}}. If $\gamma<-1$, it becomes impossible to define the limit $\eta\rightarrow0^+$. Then, it becomes reasonable to assume that $\eta$ is actually lower-bounded by $\eta_*>0$, defined by a second physical IR cut-off $\kappa_*=\eta_*\kappa_d$, competiting with the IR cut-off $\kappa_d$, due to the presence of the Casimir plates. For example, as mentioned in the body of this article, as far as the Lamb effect is concerned, this second IR cut-off is the Bethe IR cut-off $\kappa^*>\kappa_d$.

Theorem {\bf\ref{subtheorem_2a}} gives the general expression of the quantity $H^{F,\mathrm{Casimir}}_{\phi,\eta_*}(L,d)$ in the limit of perfect conductors, i.e. when $\frac{\kappa_d}{\kappa_\phi}\rightarrow0$, restricting our study to $\gamma\in\Z$ and to the following particular choice for the function $f^\beta$, depending also on the IR cut-off $\kappa_*$,
  $$
f^\beta_{\xi_*}(s)=\frac A{\kappa_d^\beta(s+\xi_*)^\beta},~~\textrm{with}~~\xi_*\geq\eta_*,
  $$
where $A$ is a fixed dimensioned constant. Theorem {\bf\ref{subtheorem_2b}} gives the general expression of the quantity $H^{F,\mathrm{Casimir}}_\phi(L,d)$, under the same assumptions.
\refstepcounter{theo} \label{theorem_2}
  \begin{subtheo}~ \label{subtheorem_2a}

Assuming that $\gamma\in\Z\smallsetminus\N$, for any $r\geq1$, in the limit $\frac{\kappa_d}{\kappa_\phi}\rightarrow0$,
  $$
H^{F,\mathrm{Casimir}}_{\phi,\eta_*}(L,d)=V\frac{\kappa_d^{\gamma+2}}{\pi^3}\!\left\{\sum_{n=1}^r\frac{b_{2n}}{(2n)!}\frac{\dd^{2(n-1)}}{\dd s^{2(n-1)}}f^\beta_{\xi_*}(s)f_\gamma(s_{\eta_*})\big|_{s=0}+O\!\!\left(\frac1{\eta_*^{2r+\beta-\gamma}}\right)\right\}\!\!.
  $$
    \begin{proof}
Since $\gamma\leq-1$, let us consider Eq. ($\bigstar^F_{\phi,0,\eta_*}$). This expression should be seen as the action of the  distribution $g\chi_{\R_+}\in\mathcal S'(\R)$ on the test function $s\longmapsto f^\beta_{\xi_*}(s)f_\gamma(s_{\eta_*})\phi(\frac{\kappa_d}{\kappa_\phi}s_{\eta_*})\in\mathcal S(\R)$. However, we remark that the action on $\mathcal S(\R)$ of its derivative $(g\chi_{\R_+})'=\pi\!\left[\sum_{p\in\N\smallsetminus\{0\}}\delta_p+\frac{\delta_0}2-\chi_{\R_+\smallsetminus\N}\right]\!\in\mathcal S'(\R)$ is simpler.

Let $\Phi_{\xi_*,\eta_*}$: $s\longmapsto\Phi_{\xi_*,\eta_*}(s)=\int_s^\infty\dd t\!~f^\beta_{\xi_*}(t)f_\gamma(t_{\eta_*})\phi(\frac{\kappa_d}{\kappa_\phi}t_{\eta_*})\in\mathcal C^\infty(\R)$. The problem is that $\Phi_{\xi_*,\eta_*}\notin\mathcal S(\R)$, nevertheless $\Phi_{\xi_*,\eta_*}\in\mathcal C^\infty(\R)$ is rapidly decreasing in the neighborhood of $\infty$ and is bounded on $\R$. Since the Schwartz class $\mathcal S(\R)$ is dense this set of functions, hereafter denoted $\mathcal B_+(\R)$ for the supremum norm $\lVert\cdot\rVert_\infty^\R$, it is common to extend the action of a particular element of $\mathcal S'(\R)$ on $\mathcal S(\R)$ to $\mathcal B_+(\R)$. This can be done here, since the distribution $(g\chi_{\R_+})'$ actually only involves positive arguments of $\Phi_{\xi_*,\eta_*}$.\footnote{As the vector space $\mathcal S'(\mathbb R)$ may be thought as the topological dual space of the vector space $\mathcal S(\mathbb R)$, there exists a classical way, based on the underlying weak-topology and density properties in Banach spaces, to extend the action of distributions.

Let be $\Psi\in\mathcal B_+(\mathbb R)$. Since $\mathcal S(\mathbb R)$ is a dense subset of $\mathcal B_+(\mathbb R)$ for the norm $\lVert\cdot\rVert_\infty^\mathbb R$, $\exists(\Psi_{\!N})_N\in\mathcal S(\mathbb R)^\mathbb N$, a so-called regularization sequence, such that $\Psi_{\!N}\xrightarrow[N\rightarrow\infty]{\lVert\cdot\rVert_\infty^\mathbb R}\Psi$. Then, by the dominated convergence theorem,
  $$
\int_\R\!\!\dd s(g\chi_{\mathbb R_+})'(s)\Psi_{\!N}(s)=\sum_{p\in\N\smallsetminus\{0\}}\Psi_{\!N}(p)+\frac{\Psi_{\!N}(0)}2-\int_{\R_+}\!\!\dd s\!~\Psi_{\!N}(s)\xrightarrow[N\rightarrow\infty]{}\sum_{p\in\N\smallsetminus\{0\}}\Psi(p)+\frac{\Psi(0)}2-\int_{\R_+}\!\!\dd s\!~\Psi(s).
  $$
As this regularized limit is well defined, this allows to extend the action of the distribution $g\chi_{\R_+}\in\mathcal S'(\R)$ to any element of $\mathcal B_+(\R)$, by setting
  $$
\int_\R\!\!\dd s(g\chi_{\mathbb R_+})(s)\Psi(s)=\sum_{p\in\N\smallsetminus\{0\}}\Psi(p)+\frac{\Psi(0)}2-\int_{\R_+}\!\!\dd s\!~\Psi(s).
  $$} Then, the action of $(g\chi_{\R_+})'\in\mathcal S'(\R)$ on $\Phi_{\xi_*,\eta_*}\in\mathcal B_+(\R)$ has a meaning, and we get
  $$
H^{F,\mathrm{Casimir}}_{\phi,\eta}(L,d)=V\frac{\kappa_d^{\gamma+2}}{\pi^3}\!\left[\sum_{p\in\N\smallsetminus\{0\}}\Phi_{\xi_*,\eta_*}(p)+\frac{\Phi_{\xi_*,\eta_*}(0)}2-\int_{\R_+}\!\!\dd s\!~\Phi_{\xi_*,\eta_*}(s)\right]\!.
  $$
Since $\Phi_{\xi_*,\eta_*}\in\mathcal C^\infty(\R)$, the Euler-Maclaurin formula can be applied to the previous expression for any $r\geq1$. Furthermore $\Phi^{(2n-1)}_{\xi_*,\eta_*}(0)=-\frac{\dd^{2(n-1)}}{\dd s^{2(n-1)}}f^\beta_{\xi_*}(s)f_\gamma(s_{\eta_*})\big|_{s=0}+O\!\!\left(\frac{\kappa_d}{\kappa_\phi}\right)$, for any $n\geq1$. In this case, by performing a Taylor expansion of the function  $t\longmapsto f^\beta_{\xi_*}(t)f_\gamma(t_{\eta_*})\in\mathcal C^\infty(\R_+)$, we obtain $\Phi^{(2n-1)}_{\xi_*,\eta_*}(0)=O\!\!\left(\frac1{\eta_*^{2(n-1)+\beta-\gamma}}\right)$. Moreover, by two integrations by parts, the remainder has the following upper bound
  \begin{align*}
\left|\int_{\R_+}\dd t~\!\frac{\widetilde B_{2r}(t)}{(2r)!}\Phi_{\xi_*,\eta_*}^{(2r)}(t)\right|&=\frac{|b_{2(r+1)}|}{(2(r+1))!}\left|\Phi^{(2r+1)}_{\xi_*,\eta_*}(0)\right|+\left|\int_{\R_+}\dd t~\!\frac{\widetilde B_{2(r+1)}(t)}{(2(r+1))!}\Phi_{\xi_*,\eta_*}^{(2(r+1))}(t)\right|
\\
&\leq\frac{|b_{2(r+1)}|}{(2(r+1))!}\!\int_{\R+}\dd t\left|\frac{\dd^{2r+1}}{\dd t^{2r+1}}f^\beta_{\xi_*}(t)f_\gamma(t_{\eta_*})\phi\!\left(\frac{\kappa_d}{\kappa_\phi}s_{\eta_*}\right)\right|\!+O\!\!\left(\frac1{\eta_*^{2r+\beta-\gamma}}\right)
\\
&=\frac{|b_{2(r+1)}|}{(2(r+1))!}C_r(\xi_*,\eta_*)+O\!\!\left(\frac1{\eta_*^{2r+\beta-\gamma}}\right)\!.
  \end{align*}

Let us study the behavior of the dimensioned constant $C_r(\xi_*,\eta_*)$, which is given by
  $$
C_r(\xi_*,\eta_*)=\int_{\R_+}\dd t\left|\frac{\dd^{2r+1}}{\dd t^{2r+1}}f^\beta_{\xi_*}(t)f_\gamma(t_{\eta_*})\phi\!\left(\frac{\kappa_d}{\kappa_\phi}t_{\eta_*}\right)\right|\!.
  $$
Since $t\longmapsto f^\beta_{\xi_*}(t)f_\gamma(t_{\eta_*})\in\mathcal C^\infty(\R_+)$ and $f^\beta_{\xi_*}(t)f_\gamma(t_{\eta_*})\propto\frac{t^\gamma_{\eta_*}}{(t+\xi_*)^\beta}$, for any $t\geq0$, by direct calculation, we can prove that $\frac{\dd^{2r+1}}{\dd t^{2r+1}}f^\beta_{\xi_*}(t)f_\gamma(t_{\eta_*})=O(\frac1{t^{2r+\beta-\gamma+1}})$ in the neighborhood of $\infty$. However because $(2r+\beta-\gamma+1)\geq2(r+1)\geq4$, i.e. $t\longmapsto\frac{\dd^{2r+1}}{\dd t^{2r+1}}f_\gamma(t_{\eta_*})\in\mathcal L^1(\R_+,\dd t)$. From the dominated convergence theorem and $\phi\in\mathcal S(\R)$, we get in the limit of $\frac{\kappa_d}{\kappa_\phi}\rightarrow0$
  \begin{align*}
&\int_{\R_+}\dd t\left|\frac{\dd^{2r+1}}{\dd t^{2r+1}}f^\beta_{\xi_*}(t)f_\gamma(t_{\eta_*})\phi\!\left(\frac{\kappa_d}{\kappa_\phi}t_{\eta_*}\right)\right|
\\
\leq&\!~\int_{\R_+}\dd t\left|\frac{\dd^{2r}}{\dd t^{2r}}\phi\!\left(\frac{\kappa_d}{\kappa_\phi}t_{\eta_*}\right)\frac\dd{\dd t}f^\beta_{\xi_*}(t)f_\gamma(t_{\eta_*})\right|+\frac{\kappa_d}{\kappa_\phi}\eta_*\int_{\R_+}\dd t\left|\frac{\dd^{2r}}{\dd t^{2r}}tf^\beta_{\xi_*}(t)f_{\gamma-1}(t_{\eta_*})\phi'\!\left(\frac{\kappa_d}{\kappa_\phi}t_{\eta_*}\right)\right|
\\
=&\int_{\R_+}\dd t\left|\frac{\dd^{2r}}{\dd t^{2r}}\phi\!\left(\frac{\kappa_d}{\kappa_\phi}t_{\eta_*}\right)\frac\dd{\dd t}f^\beta_{\xi_*}(t)f_\gamma(t_{\eta_*})\right|+O\!\!\left(\frac{\kappa_d}{\kappa_\phi}\right)\!
\\
=&\int_{\R_+}\dd t\left|\phi\!\left(\frac{\kappa_d}{\kappa_\phi}t_{\eta_*}\right)\frac{\dd^{2r+1}}{\dd t^{2r+1}}f^\beta_{\xi_*}(t)f_\gamma(t_{\eta_*})\right|+O\!\!\left(\frac{\kappa_d}{\kappa_\phi}\right)\!
\\
=&\int_{\R_+}\dd t\left|\frac{\dd^{2r+1}}{\dd t^{2r+1}}f^\beta_{\xi_*}(t)f_\gamma(t_{\eta_*})\right|\!+o(1)=\frac1{\eta_*^{2r+\beta-\gamma}}\int_{\R_+}\dd t\left|\frac{\dd^{2r+1}}{\dd t^{2r+1}}f^\beta_{\xi_*\eta_*^{-1}}(t)f_\gamma(t_1)\right|\!+o(1),
  \end{align*}
where we use $\eta_*^\beta f^\beta_{\xi_*}(\eta_* t)=f^\beta_{\xi_*\eta_*^{-1}}(t)$, for any $t\geq0$.

First if $(\beta-\gamma)=1$, {\it i.e.} if $\beta=0$ and $\gamma=-1$, the proof of the theorem ends here, since the integral $\int_{\R_+}\dd t\left|\frac{\dd^{2r+1}}{\dd t^{2r+1}}f_{-1}(t_1)\right|$ is finite, independently from $\eta_*$, for any $r\in\N\smallsetminus\{0\}$, then $C_r(\xi_*,\eta_*)=O\!\!\left(\frac1{\eta_*^{2r+1}}\right)\!$.

On the other hand, since $(\beta-\gamma)<-1$, for any $n\in\ldc0,2r+1\rdc$ and any $t\geq0$, we have
  $$
\frac{\dd^n}{\dd t^n}f^\beta_{\xi_*\eta_*^{-1}}(t)=(-1)^n\frac{\Gamma(\beta+n)}{\Gamma(\beta)}\frac{f^\beta_{\xi_*\eta_*^{-1}}(t)}{(t+\frac{\xi_*}{\eta_*})^n}~~~~\Rightarrow~~~~\left|\frac{\dd^n}{\dd t^n}f^\beta_{\xi_*\eta_*^{-1}}(t)\right|\leq\frac{\Gamma(\beta+2r+1)}{\Gamma(\beta)}f^\beta_1(t).
  $$
Therefore, it is straightforward to see that
  $$
C_r(\xi_*,\eta_*)\leq\frac{2^{2r}}{\eta_*^{2r+\beta-\gamma}}\frac{\Gamma(\beta+2r+1)}{\Gamma(\beta)}\!~\sum_{n=0}^{2r+1}\int_{\R_+}\dd t\left|f^\beta_1(t)\frac{\dd^n}{\dd t^n}f_\gamma(t_1)\right|\!.
  $$
Finally, reasoning as above, since $(\beta-\gamma)>1$, the integral $\int_{\R_+}\dd t\left|f^\beta_1(t)\frac{\dd^n}{\dd t^n}f_\gamma(t_1)\right|$ is finite, independently from $\eta_*$, for any $n\in\ldc0,2r+1\rdc$, then $C_r(\xi_*,\eta_*)=O\!\!\left(\frac1{\eta_*^{2r+\beta-\gamma}}\right)\!$, which yields the expected result, in the limit $\frac{\kappa_d}{\kappa_\phi}\rightarrow0$.
    \end{proof}
  \end{subtheo}
  \begin{subtheo}~ \label{subtheorem_2b}

Assuming that $\gamma\in\N$, for any $r\geq\left[\frac\gamma2\right]+2$, in the limit $\frac{\kappa_d}{\kappa_\phi}\rightarrow0$, then
  $$
H^{F,\mathrm{Casimir}}_\phi(L,d)=V\frac{\kappa_d^{\gamma+2}}{\pi^3}\!\left\{\sum_{n=1}^r\frac{b_{2n}}{(2n)!}\frac{\dd^{2(n-1)}}{\dd s^{2(n-1)}}f^\beta_{\xi_*}(s)f_\gamma(s)\big|_{s=0}+O\!\!\left(\frac{\kappa_d}{\kappa_\phi}\right)\right\}\!\!.
  $$
    \begin{proof}
Part of the proof of theorem {\bf\ref{subtheorem_2a}} can be adapted here, the difference being that we start with Eq. ($\bigstar^F_{\phi,0,0}$) instead of Eq. ($\bigstar^F_{\phi,0,\eta_*}$). We accordingly replace the function $\Phi_{\xi_*,\eta_*}$ by the function $\Phi_{\xi_*,0}$. Let us give the estimation of the remainder of the Euler-Maclaurin formula written for any $r\geq\left[\frac\gamma2\right]+2$
  $$
\left|\int_{\R_+}\dd t~\!\frac{\widetilde B_{2r}(t)}{(2r)!}\Phi_{\xi_*,0}^{(2r)}(t)\right|\leq\frac{|b_{2r}|}{(2r)!}\!\int_{\R+}\dd s\left|\frac{\dd^{2r-1}}{\dd s^{2r-1}}f^\beta_{\xi_*}(s)f_\gamma(s)\phi\!\left(\frac{\kappa_d}{\kappa_\phi}s\right)\right|=\frac{|b_{2r}|}{(2r)!}C_r(\xi_*).
  $$
The behavior of the dimensioned constant $C_r(\xi_*)$ is given by
  \begin{align*}
C_r(\xi_*)&=\int_{\R_+}\dd s\left|\frac{\dd^{2r-1}}{\dd s^{2r-1}}f^\beta_{\xi_*}(s)f_\gamma(s)\phi\!\left(\frac{\kappa_d}{\kappa_\phi}s\right)\right|\!
\\
&=\left(\frac{\kappa_d}{\kappa_\phi}\right)^{\!\!\!2(r-1)+\beta-\gamma}\int_{\R_+}\dd s\left|\frac{\dd^{2r-1}}{\dd s^{2r-1}}f^\beta_{\frac{\kappa_d}{\kappa_\phi}\xi_*}(s)f_\gamma(s)\phi(s)\right|
\\
&\leq2^{2r-1}\frac{\ds\max_{n\in\ldc0,2r-1\rdc}\Gamma(\beta+n)}{\Gamma(\beta)\xi_*^\beta}\left(\frac{\kappa_d}{\kappa_\phi}\right)^{\!\!\!2(r-1)-\gamma}~\sum_{n=0}^{2r-1}\int_{\R_+}\dd s\left|\frac{\dd^n}{\dd s^n}f_\gamma(s)\phi(s)\right|\!.
  \end{align*}
By the same reasoning as above, the integral $\int_{\R_+}\dd s\left|\frac{\dd^n}{\dd s^n}f_\gamma(s)\phi(s)\right|$ is finite, independently from $\kappa_\phi$, for any $n\in\ldc0,2r-1\rdc$. Since $2(r-1)-\gamma\geq1$, for any $r\geq\left[\frac\gamma2\right]+2$, the reminder now satisfies
  $$
\left|\int_{\R_+}\dd t~\!\frac{\widetilde B_{2r}(t)}{(2r)!}\Phi_{\xi_*,\eta_*}^{(2r)}(t)\right|=O\!\!\left(\frac{\kappa_d}{\kappa_\phi}\right)\!,
  $$
which yields the expected result, in the limit $\frac{\kappa_d}{\kappa_\phi}\rightarrow0$.
    \end{proof}
  \end{subtheo}
%
%
%
\subsection{Casimir energy} \label{appendix_B3}
Here, we are interested in the computation of the standard Casimir effect using the formalism developed in appendix {\bf\ref{appendix_B}}. This consists in the determination of the contribution to the electromagnetic energy given by Eq. (\ref{E_phi(L,d)_sum}) due to the Casimir plates, {\it i.e.} the term depending on $g$, which is the correction to the density of states in vacuum
  $$
E_\mathrm{Casimir}^\phi(L,d)=\frac\pi2\frac{L^2}{d^3}\int_\R\!\dd s\!~s^2\!~(g\chi_{\R_+})(s)\phi\!\left(\frac{\kappa_d}{\kappa_\phi}s\right)\!,
  $$
where we have directly identified $F(k)=\frac k2$, and deduce that $f^\beta=\frac12$, $f_\gamma(s)=s^2$, $\beta=0$ and $\gamma=2$ before  applying theorem {\bf\ref{subtheorem_2b}}. From this result written for any $r\geq3$, we obtain the correct Casimir energy in the presence of the Casimir device
  $$
E^\phi_\mathrm{Casimir}(L,d)=-\frac{\pi^2}{720}\frac{L^2}{d^3}+O\!\!\left(\frac{\kappa_d}{\kappa_\phi}\right)\!\xrightarrow[\frac{\kappa_d}{\kappa_\phi}\rightarrow0]{}E_\mathrm{Casimir}(L,d).
  $$

\end{document}